\documentclass[12pt,showpacs,amsmath,amssymb]{revtex4}
\usepackage{bm}
\usepackage{dcolumn}
\usepackage{graphicx}
\usepackage{epstopdf}
\usepackage{color}
\usepackage{epsf}
\usepackage{epsfig}

\begin{document}

\title{Gravitomagnetic Jets}

\author{C. Chicone}

\affiliation{Department of Mathematics and Department of Physics and Astronomy, University of Missouri, Columbia,
Missouri 65211, USA }

\author{B. Mashhoon}

\affiliation{Department of Physics and Astronomy, University of Missouri,
Columbia, Missouri 65211, USA}

\begin{abstract}We present a family of dynamic rotating cylindrically symmetric  Ricci-flat  gravitational fields  whose  geodesic motions have the structure of  gravitomagnetic jets. These correspond to helical motions of free test particles up and down parallel to the axis of cylindrical symmetry and are reminiscent of the motion of test charges in a magnetic field. The speed of a test particle in a gravitomagnetic jet asymptotically approaches the speed of light. Moreover, numerical evidence suggests that jets are attractors.  The possible implications of our results for the role of gravitomagnetism in the formation of astrophysical jets are briefly discussed. 
\end{abstract}

\maketitle
\section{introduction}
A constant uniform \emph{magnetic} field configuration in an inertial frame of reference has cylindrical symmetry; therefore, the motion of a test charge in this field is such that the particle's momentum and angular momentum in the direction of the field are constants of the motion. The particle in general moves with constant speed on a helix whose axis is along the field direction; moreover, the radius and step of the helical path are constants as well. The sense of helical motion about the direction of the magnetic field is positive (negative) for a test particle of negative (positive) electric charge. If the initial velocity of the particle is normal to the direction of the magnetic field, then the particle simply moves along a circle in the plane perpendicular to the field. 

The purpose of this paper is to study the analogous situation for the motion of free test particles in a \emph{gravitomagnetic} field. The nonlinearity of this field implies that only a rough similarity may be anticipated. We investigate geodesic  motion in a rotating dynamic spacetime region that is a Ricci-flat solution of Einstein's equations with cylindrical symmetry~\cite{1}. Though this time-dependent gravitational case is considerably more complicated than the magnetic case, we find qualitatively similar phenomena. In fact, the helical motions up and down parallel to the axis of symmetry are reminiscent of the double-jet structure of certain high-energy astrophysical sources. 

Quasars and active galactic nuclei generally exhibit distinct relativistic outflows. These jets are conjectured to originate from massive rotating black holes surrounded by accretion disks; the outflows are focused beams of relativistic particles that proceed up and down along the rotation axis of the black hole (see, for instance,~\cite{n2}). Similar phenomena have been observed in other high-energy astrophysical sources such as the Galactic X-ray binary systems. While our results suggest a mechanism for astrophysical jet formation, the more complicated physical process involves general relativistic MHD~\cite{2,f}. The present work---together with previous efforts~\cite{3,4,5,6,7,9}---contributes to the purely gravitational aspects of this fundamental problem in astrophysics. 

For a subclass of the Ricci-flat solutions under consideration, we show that the geodesic equations have families of special exact solutions that we call gravitomagnetic jets.  More precisely, a gravitomagnetic jet is a set of special geodesics. These  generally exhibit helical motions about the axis of  cylindrical symmetry and their union is a non-compact connected invariant manifold that attracts all nearby geodesics. While we highlight features of these jets in the source-free cylindrical spacetime region of interest and provide strong (numerical) evidence that these families are  attractors, the interesting question of the nature of the external matter currents that could generate such a gravitational field remains beyond the scope of our present investigation. 
 
According to general relativity, a rotating mass generates a relativistic, and hence non-Newtonian, gravitomagnetic field that is due to mass current. The exterior gravitomagnetic field of the Earth has recently been \emph{directly} measured via Gravity Probe B (GP-B)~\cite{13}. On the theoretical side, solutions of Einstein's equations in the case of axial symmetry have received attention for a long time (see Ch. VIII of~\cite{16} and references cited therein). In particular, rotating solutions with cylindrical symmetry have been investigated by a number of authors (see~\cite{14,11n,15} and references therein).

Previous interesting work on exact cylindrically symmetric gravitational fields in connection with the origin and structure of astrophysical jets has mainly involved the study of geodesics in the interior of time-independent rigidly rotating dust cylinders~\cite{13n}. The behavior of free test particles in this case is directly influenced by the gravitational attraction of the rotating dust particles; to avoid this circumstance, we concentrate here on certain source-free gravitational fields that happen to depend exponentially upon time. These Ricci-flat fields are perhaps more representative of the strongly time-dependent near-zone exteriors of accreting and growing gravitationally collapsed configurations where astrophysical jets are expected to originate.
The temporal variation of the gravitational fields considered in our work leads to a significant and surprising feature of the special exact solutions of the geodesic equations. The speeds of test particles in gravitomagnetic jets start from values that are always above a certain minimum speed, rapidly increase along their paths and asymptotically approach the speed of light. The minimum speed represents the speed of circular motion perpendicular to the axis of rotation.  The range of this minimum speed turns out to be from zero up to about $0.63\, c$. 

Is the main general relativistic problem of high-energy astrophysical jets solved in our paper? Our gravitational model is certainly too elementary to be an adequate representation of the complex physical situation. Nevertheless, in our simple model free test particles appear to be exponentially accelerated to almost the speed of light resulting in ultrarelativistic jet streams parallel and antiparallel to the axis of rotation.

The plan of the paper is as follows. The class of spacetime metrics that we study in this paper is described in section II. Each metric in this class involves a function $X (r)$ that is a solution of an ordinary differential equation and must be so chosen as to render our system of cylindrical coordinates admissible in the spacetime region of interest. The procedure for the determination of an appropriate $X (r)$  is described in section III. The rotational aspects of the resulting gravitational field are discussed in section IV.  Section V treats the motion of free test particles and null rays in the spacetime region of interest. This section also contains a discussion of the special analytic solutions of the geodesic equations that form  gravitomagnetic jets. Numerical evidence that gravitomagnetic jets are indeed attractors is presented in section VI. Section VII contains a discussion of our results. For clarity of presentation, some of the detailed calculations and mathematical arguments are relegated to the appendices.  Specifically, Appendix~\ref{appen:A} contains useful formulas related to the principal dynamic spacetime metric under consideration in this paper. The proofs of the main mathematical results regarding the solutions of a certain highly nonlinear ordinary differential equation are given in Appendix~\ref{appen:B}, where, due to the nature of the subject matter, some of the notation employed is independent of the rest of the paper. Appendix~\ref{appen:C} treats the geodesics of the time-reversed metric.

\section{Spacetime Metric}

In a study devoted to gravitational radiation~\cite{1}, a solution of the source-free gravitational field equations was described and partially interpreted using rotating cylindrical gravitational waves. The present paper is about the physical interpretation of a variant of this solution in a different physical domain. 

Consider the spacetime metric given in Eq.~(35) of~\cite{1}  
\begin{equation}\label{eq:1}
-ds^2=e^{-\hat{t}} \frac{X_r}{X}(-X^2 d\hat{t}^2+\frac{1}{r^3} dr^2) +\frac{e^{-\hat{t}}}{\hat{\ell} ^2 r}(\hat{\ell} X d\hat{t}+d\Phi)^2+e^{\hat{t}} d\hat{Z}^2
\end{equation}
in $(\hat{t}, r,\Phi, \hat{Z})$ coordinates. Here we employ the notation and conventions of Ref.~\cite{1}, so that  the speed of light in vacuum is unity ($c=1$) and the metric signature is $+2$. Moreover, $X_r=dX/dr$, $\hat{\ell}$ is a constant and $(r,\Phi, \hat{Z})$ are standard circular cylindrical coordinates. The function $X$ is a solution of the differential equation (cf. Eq.~(31) of~\cite{1}) 
\begin{equation}\label{eq:fode}
r^2 X^2 \frac{d^2X}{dr^2}+\frac{dX}{dr}=0,
\end{equation}
which is a special case of the generalized Emden-Fowler equation of the type
$y''=\alpha q(x) y^n y'^m$ with $y'=dy/dx$, $m=1$, $n=-2$, $\alpha=-1$ and $q(x)=x^{-2}$ (see~\cite{14n}).  We transform it to the Lotka-Volterra system~\eqref{eq:hsi3} in Appendix~\ref{appen:B}. We do not know an explicit general solution of Eq.~\eqref{eq:fode}.  For the only explicit solutions known to exist, we note that $X=$ constant is unacceptable as the 4D metric~\eqref{eq:1} would then degenerate into a 3D spacetime, and the  solutions 
\begin{equation}\label{eq:3}
X=\pm \big(\frac{3}{2}r\big)^{-1/2}
\end{equation}
correspond to the special free rotating gravitational waves investigated in detail in~\cite{1} and the references cited therein. 

About fifteen years ago, searching for a general description of rotating
cylindrical gravitational waves, one of us (BM) found the Ricci-flat metric~\eqref{eq:1}, where $X$ is a solution of Eq.~\eqref{eq:fode} once $R_{\mu\nu}=0$.  Among the solutions of Eq.~\eqref{eq:fode},
there is a class for which $X^2$ is a monotonically decreasing function of $r$
as well as a class where $X^2$ is monotonically increasing. The former
class of solutions, which includes Eq.~\eqref{eq:3} as a special case, was
investigated in Ref.~\cite{1} and related to rotating
gravitational waves. The physical interpretation of the latter class of
solutions is taken up in the present work.

Let us note that if $X$ is a solution of Eq.~\eqref{eq:fode}, then so is $-X$; therefore, it is possible to assume $\hat{\ell}>0$ in Eq.~\eqref{eq:1} with no loss of generality. We define a lengthscale $\lambda$ such that 
$
\hat{\ell}:=\lambda^{-1/2}
$. 
Only positive square roots are considered throughout. 
Let us now introduce the \emph{dimensionless} quantities $\tilde r$, $\tilde{X}$ and $\tilde{z}$ such that 
\begin{equation}\label{eq:5}
\tilde r=\lambda r,\qquad \tilde X=\lambda^{-1/2}X,\qquad \tilde z=\lambda^{-1}\hat{Z}.
\end{equation} 
Moreover, $\hat{t}$ in Eq.~\eqref{eq:1} is dimensionless as well; therefore, we assume that the physical time coordinate is given by $\lambda'\hat t$, where $\lambda'$ is in general a different arbitrary constant lengthscale. Under the scale transformation $(r,X)\mapsto (\tilde r,\tilde X)$, Eq.~\eqref{eq:fode} remains invariant. Furthermore, with 
$
\tilde{s}=\lambda^{-1} s
$ 
the spacetime interval~\eqref{eq:1} essentially remains invariant as well; that is, dropping all the tildes and working only with \emph{dimensionless} quantities, the spacetime metric takes the form 
\begin{equation}\label{eq:7}
-ds^2=e^{-{t}} \frac{X_r}{X}(-X^2 d{t}^2+\frac{1}{r^3} dr^2) +\frac{e^{-{t}}}{ r}( X d{t}+d\phi)^2+e^{{t}} dz^2,
\end{equation}
where $\hat t=t$ and $\Phi=\phi$. Starting from dimensionless $(t, r,\phi, z)$ coordinates, we can always return to regular coordinates by choosing arbitrary lengthscales $\lambda$ and $\lambda'$; then, the physical coordinates are $(\lambda' t,\lambda^{-1} r, \phi, \lambda z)$ and the other physical quantities in the metric are $\lambda^{1/2} X$ and $\lambda s$.

For the physical interpretation proposed in this work, the spacetime metric is obtained from Eq.~\eqref{eq:7} by $t\mapsto -t$. Henceforth, we will deal with dimensionless quantities and the spacetime metric
 \begin{equation}\label{eq:8}
-ds^2=e^{t} \frac{X_r}{X}(-X^2 d{t}^2+\frac{1}{r^3} dr^2) +\frac{e^{t}}{ r}(- X d{t}+d\phi)^2+e^{-t} dz^2,
\end{equation}
where $X$ is a solution of Eq.~\eqref{eq:fode} specified in the next section, and we will show that under certain conditions geodesics of metric~\eqref{eq:8} allow gravitomagnetic jets. Thus the main focus of the following sections and Appendices~\ref{appen:A} and~\ref{appen:B} is on metric~\eqref{eq:8}; we return to the time-reversed case---namely, metric~\eqref{eq:7}---in Appendix~\ref{appen:C}. 

As shown in~\cite{1}, Eq.~\eqref{eq:8} represents an algebraically general Ricci-flat solution of type I in the Petrov classification. It admits two commuting spacelike Killing vector fields $\partial_z$ and $\partial_\phi$ associated with the cylindrical symmetry of the gravitational field. The corresponding two-parameter isometry group is not orthogonally transitive. The invariant magnitude of the hypersurface-orthogonal $\partial_z$ is given by $\exp(-\frac{1}{2}t)$, while for $\partial_\phi$, which is not hypersurface-orthogonal, the invariant magnitude is $r^{-1/2}\exp(\frac{1}{2}t)$. Thus $t$ and $r$ can be invariantly defined in this way.  The $t=$ constant hypersurfaces are always spacelike. We interpret $t$ as the time coordinate in this paper; therefore, at a given time $t$,  
$
\rho=r^{-1/2}
$
is, up to a constant factor, an appropriate \emph{radial} coordinate. Using this quantity, metric~\eqref{eq:8} takes the form
 \begin{equation}\label{eq:10}
-ds^2=-\frac{1}{2}e^{t}\rho^3 \frac{X_\rho}{X}(-X^2 d{t}^2+4  d\rho^2) +e^{t}\rho^2(- X d{t}+d\phi)^2+e^{-t} dz^2,
\end{equation}
so that the circumference of a circle orthogonal to the axis of cylindrical symmetry is $2\pi\rho^*$, where $\rho^*=\rho \exp(\frac{1}{2}t)$. The symmetry axis is thus defined by $\rho=0$ (or $r=\infty$). Despite its shortcoming as a proper radial coordinate, we nevertheless use $r$ extensively in this paper to simplify formulas. 

We define the proper radial distance           $R^*=\exp(t/2) R$                        via Eq.~\eqref{eq:10} such that for the proper radial distance from the axis to a  point with  $\rho=\rho_0$, $R=R_0$                   is given by
\begin{equation}\label{eq:nn11}
R_0=\int_0^{\rho_0}\big(-2 \rho^3 \frac{X_\rho}{X}\big)^{1/2} \,d\rho=\int_{r_0}^\infty\big(\frac{X_r}{r^3 X}\big)^{1/2}\,dr,
\end{equation}
where $r_0=1/\rho_0^2$.

The coordinates $x^\mu=(t,r,\phi,z)$ are assumed to be \emph{admissible} in the spacetime domain under consideration here. Thus, the gravitational potentials must be sufficiently smooth functions of these coordinates; moreover, the Lichnerowicz admissibility conditions~\cite{8} require that the \emph{principal minors} of the metric tensor and its inverse be negative for our choice of metric signature. For a symmetric $n\times n$ matrix $M$, the principal minors are given by 
\begin{equation}\label{eq:i}
\det \left[
\begin{array}{ccc}
M_{11} & \cdots & M_{1k}\\
\vdots & &\vdots\\
M_{k1} & \cdots & M_{kk}
\end{array}
\right]
\end{equation} 
for $k=1,\ldots, n$.
These admissibility conditions are essentially equivalent to the inequalities
\begin{equation}\label{eq:11a}
X(r X_r-X)>0, \qquad XX_r>0,
\end{equation}
for all values of $r$ in the domain of the solution $X$. This fact 
follows from a detailed examination of the spacetime metric tensor $g_{\mu\nu}$, with $-ds^2=g_{\mu\nu} dx^\mu dx^\nu$ defined in display~\eqref{eq:8},  and its inverse given in Appendix~\ref{appen:A}; we note, in particular,  that for an admissible solution 
\begin{equation}\label{eq:11b}
-g^{tt}=\frac{e^{-t}}{X X_r}>0,\qquad (-g)^{-1/2}=\frac{r^2 e^{-t}}{|X_r|}>0. 
\end{equation}

There are four algebraically independent scalar polynomial curvature invariants in a Ricci-flat spacetime that can be represented as (see Ch.~9 of~\cite{14})
\begin{align}
\label{eq:13} \mathcal{I}_1&=R_{\mu\nu\rho\sigma}R^{\mu\nu\rho\sigma}-iR_{\mu\nu\rho\sigma}R^{*\mu\nu\rho\sigma}, \\
\label{eq:14} \mathcal{I}_2&=R_{\mu\nu\rho\sigma}R^{\rho\sigma\alpha\beta}R_{\alpha\beta}^{\hspace{.2in}\mu\nu}+iR_{\mu\nu\rho\sigma}R^{\rho\sigma\alpha\beta}{R^*}_{\alpha\beta}^{\hspace{.2in}\mu\nu}.
\end{align}
For metric~\eqref{eq:8}, $\mathcal{I}_1$ and $\mathcal{I}_2$ are both real and are given by
\begin{align}
\label{eq:15} \mathcal{I}_1&=-\frac{e^{-2t}}{r X^4X_r^2}(1-3 r X^2-2 r^2 X X_r-r^3 X^3 X_r+r^4 X^2 X_r^2), \\
\label{eq:16} \mathcal{I}_2&=-\frac{3e^{-3t}}{4r X^5X_r^3}(1-2 r X^2-2 r^2 X X_r+2r^3 X^3 X_r+r^4 X^2 X_r^2).
\end{align}
These invariants vanish in the case of  special solution~\eqref{eq:3}, but are in general nonzero. For any finite value of the coordinate time $t$, $\mathcal{I}_1$ and $\mathcal{I}_2$ depend on the particular branch of the function $X$. This will be discussed in detail in section III. 

\section{Choice of X}
\begin{figure}
\begin{center}
\psfig{file=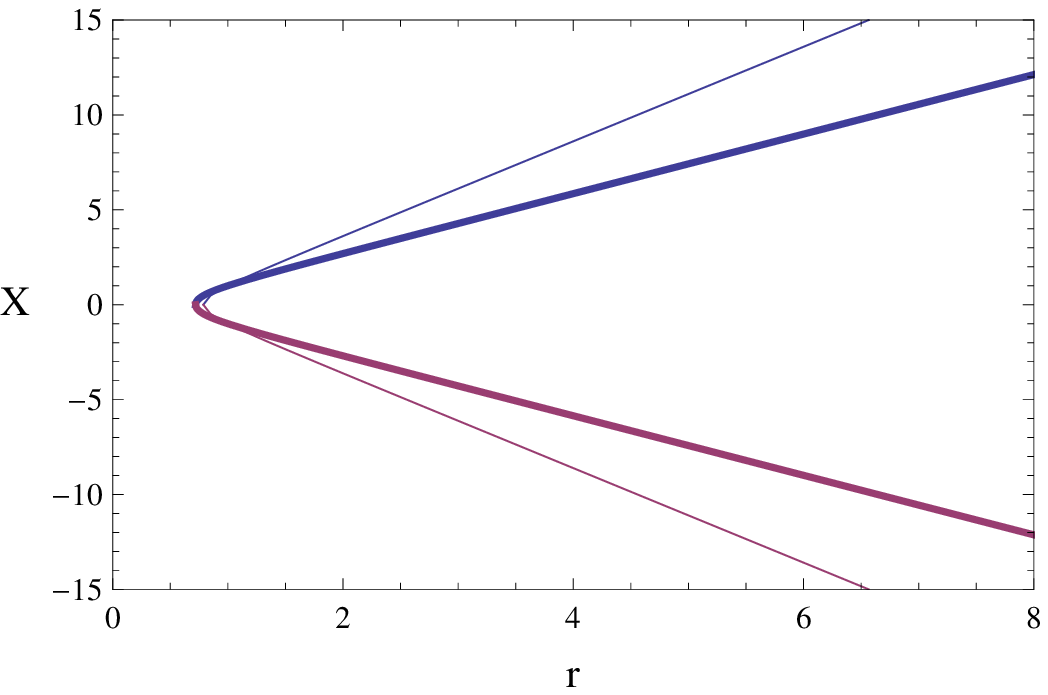, width=20pc}\\
\psfig{file=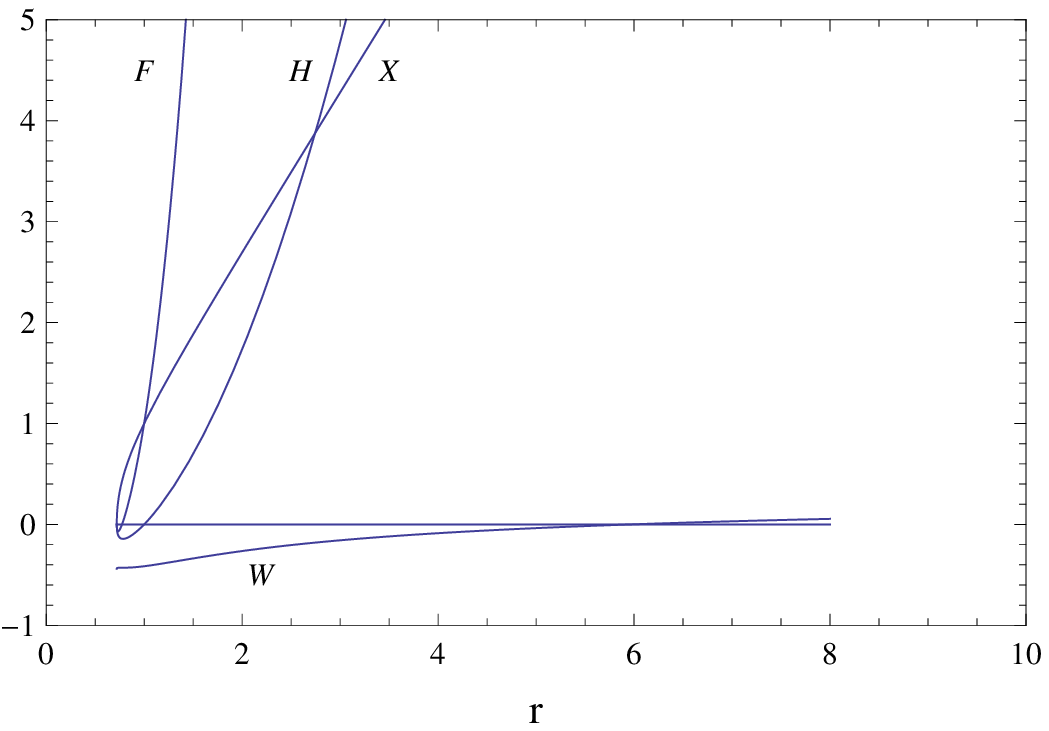, width=20pc}\\
\psfig{file=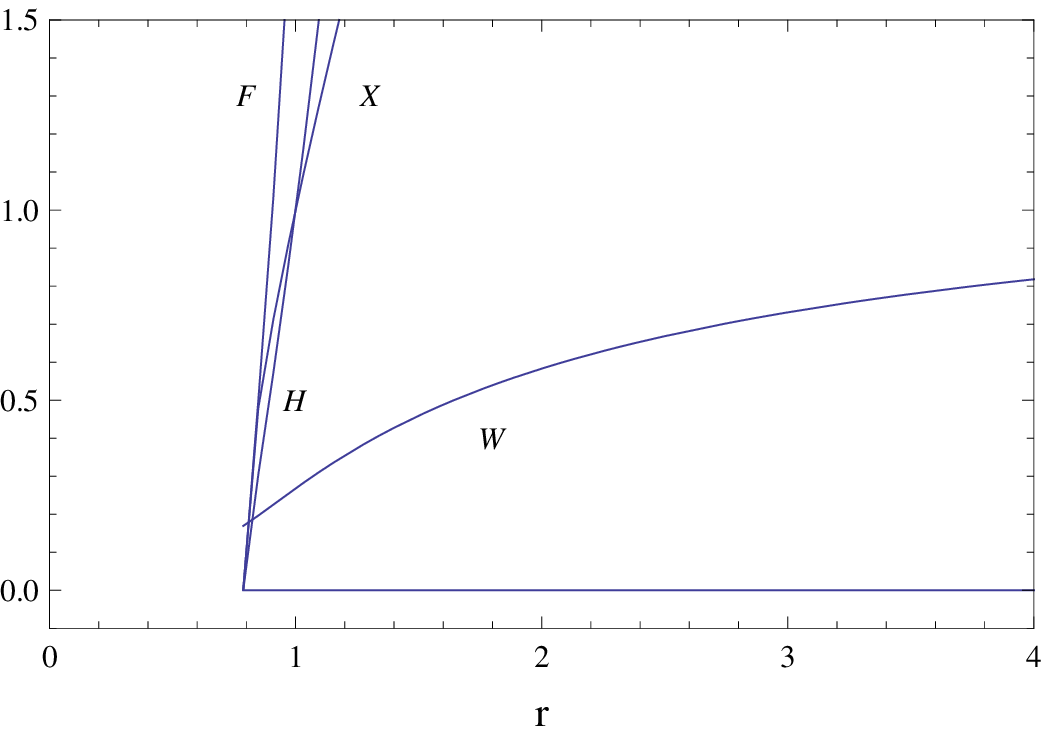, width=20pc}
\end{center}
\caption{The top panel shows computer-generated plots of admissible solutions $\pm X$ versus $r$ for the differential Eq.~\eqref{eq:fode}. The thick curve is for the initial data $X(1)=1$ and $X_r(1)=2$; the thin curve is for the data $X(1)=1$ and $X_r(1)=3$. The middle (bottom) panel depicts computer generated graphs of $F$, $H$, $W$ and  $X$ for the ``thick" (``thin") solution. The graphs of $F$, $H$ and  $X$ vanish at the left end-point $r_b$  of the domain of definition of $X$, which is  $r_b\approx 0.720$ for the thick solution that allows jets and $r_b\approx 0.787$ for the thin solution that does not allow jets.  
\label{fig:2}}
\end{figure}
The mathematical investigation needed to identify admissible solutions  $X(r)$   of Eq.~\eqref{eq:fode} that would allow the existence of jets is given in Appendix~\ref{appen:B}; here, we present the main conclusions of this analysis.

The character of metric tensor~\eqref{eq:8} depends on the choice of a solution of the differential Eq.~\eqref{eq:fode}.  We will consider admissible solutions in $(t,r,\phi,z)$ coordinates that correspond to solutions $X$ of the differential Eq.~\eqref{eq:fode} such that $X^2$ is monotonically increasing with $r$ and  
\begin{equation}\label{eq:admiss}
Q(r):=r X_r-X
\end{equation}
is such that $X Q>0$. 

Throughout this paper, we take advantage of the scale transformation already mentioned in the previous section: The differential Eq.~\eqref{eq:fode} remains invariant under the scaling $(r,X)\mapsto(\hat{r},\hat{X})$, where       for $\sigma>0$, $\hat r=\sigma r$ and  $\hat X(\hat{r})=\sigma^{-1/2} X(r)$. Using scale-invariant variables, it is then possible to reduce the second-order Eq.~\eqref{eq:fode} to equations of first order---see Appendix~\ref{appen:B}. Indeed, consideration of scale-invariant quantities leads to substantial simplifications in our work. The admissible class that we seek is a scale-invariant subclass of solutions of Eq.~\eqref{eq:fode} that are all of the form depicted for two cases in the top panel of Figure 1. 
In fact, in Appendix~\ref{appen:B} we prove that there is an open set of initial conditions that corresponds to admissible solutions each of which exists on an interval $(r_b,\infty)$,  where  $r_b>0$,  with limiting values $X(r_b)=0$ and $X_r(r_b)=\infty$ such that $X X_r=1/r_b^2$ at $r_b$, the function  $X^2$ is increasing, $X_r^2$ is decreasing and $X^2$ approaches infinity as $r$ approaches infinity. In addition, we show that 
there is an open set of initial conditions for solutions defined on $(r_b,\infty)$ with the same end-point asymptotics such that  
the scale-invariant function $F$ defined by
\begin{equation}\label{eq:F(r)}
F(r)=r^2 X(r)X_r(r)-1
\end{equation}
has a zero $r_J$  in the interior of the interval of existence $(r_b,\infty)$. The physical significance of this property 
is that it is the necessary and sufficient condition for the existence of jets; that is, if   $F$   has such a zero, then the (timelike and null) geodesic equations admit special helical solutions propagating upward and downward on an interior cylinder of radius  $r_J$. These special solutions form gravitomagnetic jets.  We also show that there is an open set of initial data corresponding to solutions that satisfy both conditions.
\begin{figure}
\centerline{\psfig{file=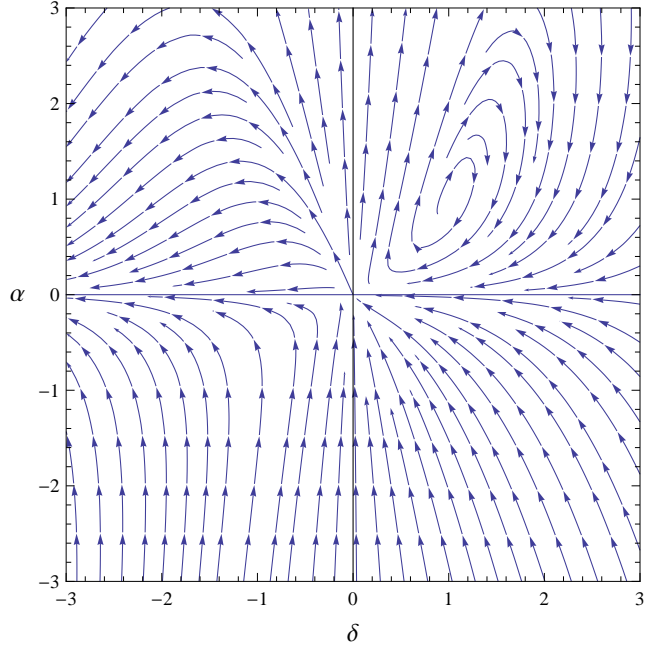, width=20pc}}
\caption{Computer generated phase portrait for system~\eqref{eq:asfo}. The rest point at  $(\delta,\alpha)=(0,0)$ is a degenerate singularity of system~\eqref{eq:asfo}, while the other rest point at $(1,1)$ is a simple singularity. This latter isolated singularity is a spiral point that corresponds to the special solution~\eqref{eq:3}. 
\label{fig:1}}
\end{figure}

To clarify the connection of the present work with the solutions treated in Ref.~\cite{1}, let us consider, as in~\cite{1},  the scale-invariant quantities 
\begin{equation}\label{eq:xideltadef}
\delta=\frac{2}{3 r X^2},\qquad \alpha=-\frac{4 X_r}{3 X^3}
\end{equation}
and note that these quantities are related by the autonomous differential equation  
\begin{equation}\label{eq:as}
\frac{d\alpha}{d \delta}=\frac{3\alpha (\alpha-\delta^2)}{2\delta (\alpha-\delta)}.
\end{equation}
By introducing a temporal parameter $\theta$, we will also consider the corresponding first-order system
\begin{equation}\label{eq:asfo}
\begin{split}
\frac{d\delta}{d \theta}&=2\delta (\alpha-\delta),\\
\frac{d\alpha}{d \theta}&=3\alpha (\alpha-\delta^2),
\end{split}
\end{equation}
whose phase portrait is depicted in Fig.~\ref{fig:1}. With the choice of $r$ as a positive radial coordinate in metric~\eqref{eq:8}, $\delta\ge 0$ and hence the first and fourth quadrants of Fig.~\ref{fig:1} are physically relevant. The solutions in the first quadrant have $\alpha>0$ or $X X_r<0$, so that $X^2$ monotonically decreases with $r$. These solutions, which were discussed in~\cite{1}, tend to the special rest point $(\delta,\alpha)=(1,1)$ that corresponds to exact solution~\eqref{eq:3}. The present paper is therefore devoted to the study of solutions in the fourth quadrant for which $\alpha<0$ and hence $X X_r>0$. Since system~\eqref{eq:asfo} is autonomous and the coordinate axes are invariant, it follows immediately that the fourth quadrant is invariant; that is, solutions that start in the fourth quadrant stay there~\cite{ccc}. 
In particular, $X(r) X_r(r)>0$ as long as such a solution exists.
If $X(r)>0$, then $X_r(r)>0$,  $X$ increases as $r$ increases, and $X_{rr}(r)$ is negative. Thus, if $X(r)>0$ for some $r>0$, this function increases as $r$ increases and is concave down. Using our symmetry, if $X(r)<0$ for some $r>0$, then $X$ decreases with $r$ and is concave up (see Fig.~\ref{fig:2}). 
Each member of this class of solutions of Eq.~\eqref{eq:fode} is asymptotic to the degenerate rest point $(\delta,\alpha)=(0,0)$ that corresponds to the axis of cylindrical symmetry, where $r=\infty$ and $X^2=\infty$. In fact, it can be shown that, with our choice of temporal variable $\theta$,  each solution is asymptotic to  $(0,0)$ (corresponding to  axis of cylindrical symmetry) in positive time and approaches a point at infinity $(\delta,\alpha)=(\infty,-\infty)$ in negative time  (corresponding to the cylindrical boundary $r=r_b>0$ and $X X_r=r_b^{-2}$).

\begin{figure}[h]
\begin{center}
\psfig{file=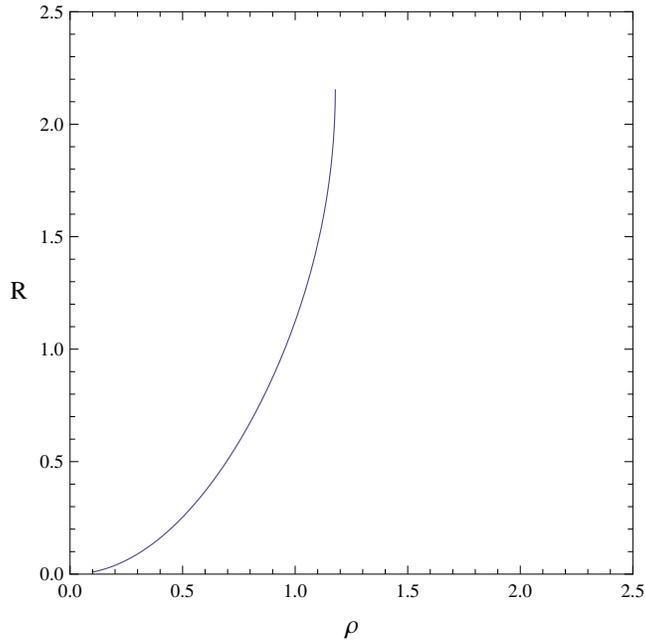, width=20pc}
\end{center}
\caption{Plot of $R$ versus $\rho$ corresponding to the spacetime defined by $X$ with  $X(1)=1$ and $X_r(1)=2$, as in the middle panel of Fig.~\ref{fig:2}. 
\label{fig:rvrho}}
\end{figure}
Consider a hypersurface such that at every instant of time $t$, the hypersurface is characterized by the cylindrical surface     $r$ = constant for $r_b<r<\infty$. The gravitomagnetic jets lie on such a hypersurface with  $r=r_J$. The normal to such a hypersurface is given by $n_\mu=(0,1,0,0)$; hence, $n_\mu n^\mu =g^{rr}=\exp(-t) r^3 X/X_r>0$  using Eq.~\eqref{A2} of Appendix A. That is, the hypersurface is always timelike, but as $r\to r_b^+$, $n_\mu n^\mu\to 0$, so that the boundary hypersurface is null.

The admissibility conditions require that we explicitly leave the inner and outer boundaries out of the domain of the solution $X$. That is, the axis of cylindrical symmetry ($r=\infty$) and the boundary cylinder ($r=r_b$) are excluded from the spacetime region of physical interest, since  $g^{tt}$ vanishes at the axis and $(-g)^{-1/2}$ vanishes at the boundary (see Eq.~\eqref{eq:11b}).

It is interesting to describe further some of the significant properties of the admissible solutions. Let us first note that the invariance of Eq.~\eqref{eq:fode} under  $X\mapsto -X$ is a scale-invariant property; \emph{henceforth, we will work exclusively with the   $X>0$               branch}. We show in Appendix~\ref{appen:B} that near the axis ($\rho\to 0$ and $r\to \infty$),
\begin{equation}\label{eq:19m}
X(r)\sim ar +a S_0-\frac{1}{6a r^2}+\frac{S_0}{6 a r^3}+O(\frac{1}{r^4}),
\end{equation}
where  $a>0$         and   $S_0<0$       are constants. This implies, among other things, that the function   $Q>0$, which transforms like  $X$      under scaling and satisfies $dQ/dr=-X_r/(r X^2)$, decreases monotonically from infinity at the boundary cylinder to  $Q(\infty)=-a S_0>0$  at the axis of symmetry. The behavior of  $X(r)$ near the boundary cylinder is more complicated (see Appendix~\ref{appen:B}); we show there that for sufficiently small $r-r_b>0$,
\begin{equation}\label{eq:20m}
r^2 X X_r = 1+ \Big(\frac{2}{r_b}\Big)^{1/2}A (r-r_b)^{1/2} + \frac{2(6+A^2)}{3 r_b} (r-r_b)+O((r-r_b)^{3/2}),
\end{equation}
where  $A$  is a constant (cf. Eq.~\eqref{eq:iB10}).
This is a particularly useful relation as it involves the behavior of $F(r)$ as $r\to r_b$. For the admissible solutions that allow jets, $A<0$.

When $A\ge 0$,  $F(r)$ is positive for $r$ in some interval whose left endpoint is $r_b$, while for $A<0$,  it is negative.  The admissible solutions that allow jets are a subset of  solutions of Eq.~\eqref{eq:fode} for which $A<0$.  Using only the leading term of the series in Eq.~\eqref{eq:20m}, we find the asymptotic differential equation
\begin{equation} \label{eq:bm25}
X(r)X_r(r)=\frac{1}{r^2}.
\end{equation}
Its solution is given by
\begin{equation} \label{eq:bm26a} X(r)=\pm \big[X(r_0)^2-\frac{2}{r}+\frac{2}{r_0}\big]^{1/2}.
\end{equation}
In the special case of the boundary cylinder where  $X(r_b)=0$ and $r>r_b$, 
\begin{equation} \label{eq:bm26}
X(r)=
\pm \big(\frac{2}{r}\big)^{1/2}\big(\frac{r}{r_b}-1\big)^{1/2}.
\end{equation}
Based on this result, we let $X(r)=\epsilon^{1/2} \chi(\epsilon)$, where $\epsilon=-1+r/r_b$; then, Eq.~\eqref{eq:fode} implies a nonlinear second-order equation for $\chi(\epsilon)$. It turns out that this equation has a unique analytic solution near $\epsilon=0$ for $A=0$, see Eq.~\eqref{eq:A15}. For $A\ne 0$, there is no analytic solution; however, Eq.~\eqref{eq:t6bb} implies that  as $r\to r_b^+$ 
\begin{equation}\label{eq:leq3}
X ( r ) = \pm \Big (\frac{2}{r_b}\Big )^{1/2} [\epsilon^{1/2} + \frac{2^{1/2}A}{3}  \epsilon + \frac{(9 + A^2)}{18}\epsilon^{3/2} + O(\epsilon^2) ]. 
\end{equation}

It follows from Eqs.~\eqref{eq:19m}--\eqref{eq:20m} that for an admissible solution the proper radial distance from the axis to the boundary---namely, $\exp(t/2)R_b$---is finite at every given finite instant of time $t$. Moreover, near the axis $\rho=0$, $R\approx \rho^2$, while $R$ has infinite slope as $\rho\to \rho_b$. Figure~\ref{fig:rvrho} illustrates $R$ versus $\rho$ for the solution given in the middle panel of Fig.~\ref{fig:2}. We note that the requirement of elementary flatness is violated near the axis of cylindrical symmetry. That is, for an infinitesimal spacelike circle around the axis, the ratio of the circumference, $2 \pi \rho \exp(t/2)$, to the proper radius, $\rho^2\exp(t/2)$, is $2\pi/\rho$ instead of $2\pi$ at a given instant of time $t$ and  diverges as $\rho\to 0$. For a thorough discussion of subtle issues regarding cylindrical symmetry and its axis, see~\cite{19} and the references cited therein. We recall that in our analysis the axis ($\rho=0$) is already excluded from the physical domain in order to satisfy the admissibility conditions.

To help distinguish the admissible solutions that do allow jets from those that do not, it is useful to consider two other functions associated with $X$:    $H(r)$   and       $W(r)$. The scale-invariant  $H$      is defined by
\begin{equation}\label{eq:21m}
H(r)=r X Q-1=F-r X^2,
\end{equation}
so that  $H(r_b)=0$, while for  $r\to\infty$, $H\sim -a^2 S_0r^2$; moreover, if $X$      allows
jets, $-1<H(r_J)<0$  by  Eq.~\eqref{eq:21m}. The function   $W$     scales as  $X$     and is defined by
\begin{equation}\label{eq:22m}
W= Q-\Big(\frac{X_r}{X}\Big)^{1/2}.
\end{equation}
Thus   $W$    is positive near the axis; in fact,  $W(\infty)=-a S_0>0$.         For the value of   $W$  near the boundary, let us first note that   $W$    can be written as
\begin{equation}\label{eq:23m}
W= \frac{HQ-X}{1+H+(r^2 X X_r)^{1/2}}
\end{equation}
and for $r\to r_b^+$ , $HQ\to  A/r_b^{1/2}$  from Eq.~\eqref{eq:20m}. Thus
$
W(r_b)= A/(2r_b^{1/2})
$.
The solutions that allow jets are among those for which   $A<0$. Figure~\ref{fig:2} illustrates, via the behavior of these functions, the difference between the solutions that allow jets and those that do not. The top panel contains the graphs of two admissible solutions: The thick (thin) curve represents a solution that allows (does not allow) jets. The middle (bottom) panel illustrates the behavior of  $F$,  $H$    and $W$     for the admissible solution that allows (does not allow) jets.
 
It is important to point out that for the admissible solutions, the curvature invariants $\mathcal{I}_1$ and $\mathcal{I}_2$ given by Eqs.~\eqref{eq:15} and~\eqref{eq:16} are indeed finite for finite values of time $t$ and radial coordinate $r\in (r_b,\infty)$.  We remark in passing that they diverge in the infinite past ($t\to -\infty$); however, this circumstance is consistent with the emergence of the universe from a singular state as in the standard models of cosmology.  Using Eq.~\eqref{eq:F(r)}, it is possible to express these scalars as
\begin{align} \label{eq:25m}
\begin{split}
\mathcal{I}_1&=-\frac{e^{-2t}}{r X^4 X^2_r}[F^2-r X^2(F+4)],\\
\mathcal{I}_2&=-\frac{3e^{-3t}}{4r X^5 X^3_r}F(F+2 r X^2).\\
\end{split}
\end{align}
Thus $\mathcal{I}_1(r_J)=4 r^4_{J}\exp(-2 t)$                                            and   $\mathcal{I}_2(r_J)=0$ for any solution that allows jets. Specifically, for the case that allows jets depicted in the middle panel of Fig.~\ref{fig:2},  $ \mathcal{I}_1$   has one zero and     $ \mathcal{I}_2$    has two zeros in the interval $(r_b,\infty)$, while for the case that does not allow jets depicted in the bottom panel of Fig.~\ref{fig:2},     $ \mathcal{I}_1$    has one zero and    $ \mathcal{I}_2$     has no zeros in the interval  $(r_b,\infty)$.  Moreover, it is straightforward to show that as the boundary cylinder is approached ($r\to r_b$),
\begin{equation}\label{eq:39n}
\mathcal{I}_1\to (4- A^2) e^{-2 t} r_b^{4}, \qquad \mathcal{I}_2\to -\frac{3A^2}{4} e^{-3t} r_b^6,
\end{equation}
while  as the axis of cylindrical symmetry is approached ($r\to \infty$), 
 \begin{equation}\label{eq:40n}
\mathcal{I}_1\to e^{-2 t} S_0 a^{-2}, \qquad \mathcal{I}_2\to -\frac{9}{4}e^{-3 t}a^{-4}. 
\end{equation}

Once $X$ has been properly chosen, we can turn to the treatment of gravitational physics in the corresponding singularity-free spacetime region. This is an open hollow cylindrical domain that expands; it has an inner boundary around the symmetry axis ($r=\infty$) and an outer boundary ($r=r_b$). We therefore consider test particles and gyroscopes in the radial interval $(r_b,\infty)$ in the rest of this paper.

The rotational aspects of the spacetimes under consideration here provide the basis for the interesting features that free test particle motion can exhibit in these gravitational fields; therefore, we now turn to the gravitomagnetic properties of admissible solutions.

\section{Gravitomagnetic Field} 

The gravitational Larmor theorem implies a local equivalence, in the linear approximation, between gravitomagnetism and rotation~\cite{11,12}. This correspondence may be employed in order to provide a definite measure of the gravitomagnetic field. It is therefore useful to study the precession of ideal test gyroscopes  that are held at rest in the spacetime region of interest; the precession frequency may then be identified with the gravitomagnetic field for the class of observers that carry the gyros along their world lines. 

To simplify matters, we  consider the class of \emph{fundamental} observers that are spatially at rest by definition; that is, $x^i$ is constant for each $i=1,2,3$ for a fundamental observer in the physical spacetime region of interest. Therefore, the four-velocity field of the fundamental observers is given  in  $(t,r,\phi,z)$ coordinates by 
\begin{equation}\label{eq:41n}
\lambda^\mu_{\hspace{0.07in} (t)}=((-g_{tt})^{-1/2},0,0,0). 
\end{equation} 
These observers' natural orthonormal tetrad frame is given by $\lambda^\mu_{ \hspace{0.07in}(\alpha)}$, where
\begin{align}
\label{eq:42a} \lambda^\mu_{ \hspace{0.07in}(r)}&=(0, (g_{rr})^{-1/2},0,0),\\
\label{eq:42b} \lambda^\mu_{ \hspace{0.07in}(\phi)}&=(s^t, 0, s^\phi,0),\\
\label{eq:42c} \lambda^\mu_{ \hspace{0.07in}(z)}&=(0, 0, 0,e^{t/2})
\end{align}
are the corresponding spatial unit directions. Here, $s^t$ and $s^\phi$ are given by
\begin{equation}\label{eq:43n}
s^t=-e^{-t/2}(X_r Q)^{-1/2}, \qquad s^\phi=e^{-t/2} \big(\frac{Q}{X_r}\big)^{1/2}. 
\end{equation}  

The fundamental observers are accelerated; their acceleration tensor $\omega_{(\alpha)(\beta)}$ is defined via
\begin{equation}\label{eq:44n}
\frac{D\lambda^\mu_{ \hspace{0.07in}(\alpha)}}{dT}=\omega_{(\alpha)}^{\hspace{0.14in}(\beta)} \lambda^\mu_{ \hspace{0.07in}(\beta)},
\end{equation}
where $T$ is the proper time along the world line $x^\mu(T)$ of a fundamental observer such that $dx^\mu/dT=\lambda^\mu_{ \hspace{0.07in}(t)}$. In analogy with electrodynamics, the antisymmetric acceleration tensor consists of an ``electric'' part $\omega_{(t)(i)}=\mathcal{A}_{(i)}$ and a ``magnetic'' part $\omega_{(i)(j)}=\epsilon_{(i)(j)(k)}\Omega^{(k)}$. The corresponding spacelike vectors are then $\mathcal{A}_\mu=\mathcal{A}_{(i)}\lambda_\mu^{ \hspace{0.07in}(i)}$ and $\Omega^\mu=\Omega^{(k)}\lambda^\mu_{\hspace{0.07in}(k)}$. It follows from a detailed calculation that in the present case
\begin{align}
\label{eq:45a}\mathcal{A}_{(r)}&=-\frac{1}{2}e^{-t/2}\Big(\frac{rX_r}{X}\Big)^{1/2}\,\frac{X_r-X Q^2}{XX_r Q},\\
\label{eq:45b}\mathcal{A}_{(\phi)}&=-\frac{1}{2}e^{-t/2}(X_r Q)^{-1/2}
\end{align}
are the nonzero tetrad components of the translational acceleration $\mathcal{A}_\mu$ of the fundamental observers. These are finite in the physical region $(r_b,\infty)$; moreover, as the symmetry axis is approached 
\begin{equation}\label{x36}
\mathcal{A}_{(r)}\to -\frac{1}{2}e^{-t/2}S_0>0,\quad 
\mathcal{A}_{(\phi)}\to -\frac{1}{2}e^{-t/2} a(- S_0)^{-1/2}<0,
\end{equation}
while near the boundary $r\to r_b^+$ one can show, using the expressions given in the previous section, that 
\begin{equation}\label{x37}
\mathcal{A}_{(r)}\to \frac{1}{2}e^{-t/2}A r_b,\quad 
\mathcal{A}_{(\phi)}\to 0.
\end{equation}
Furthermore, $\Omega^\mu$ can be obtained from 
\begin{equation}\label{eq:46n}
\Omega^{(z)}=-\frac{1}{2}e^{-t/2}(XQ)^{-1},
\end{equation}
which is the rotation frequency about the $z$ axis of the local spatial frame of the fundamental observers with respect to the local \emph{nonrotating}---that is, Fermi-Walker transported---frame. All of the other nonzero components of the acceleration tensor can be obtained from Eqs.~\eqref{eq:45a}--\eqref{eq:45b} and~\eqref{eq:46n}.

\begin{figure}
\begin{center}\psfig{file=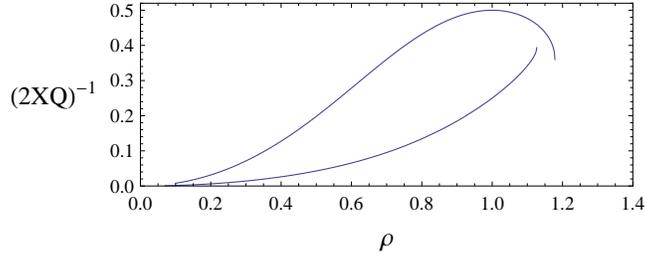, width=20pc}
\end{center}
\caption{\label{fig:4}Plots of $(2 X Q)^{-1}$ versus $\rho=r^{-1/2}$ for the two cases of $X$ given in the top panel of Fig.~\ref{fig:2}. The upper curve here corresponds to initial data $X(1)=1$ and $X_r(1)=2$; the lower curve corresponds to 
$X(1)=1$ and $X_r(1)=3$. In general, the maximum of $(2 X Q)^{-1}$ occurs where $H(r)=r XQ-1$ vanishes; this happens at $r=1$ for the upper curve.}
\end{figure}
These considerations imply that ideal test gyroscopes carried along the world lines of the fundamental observers precess, with frequency $-\Omega^{(z)}$ about the $z$ axis, with respect to the natural spatial frame of the fundamental observers.  Thus
\begin{equation}\label{eq:47n}
-\Omega^{\mu}=(0,0,0, \frac{1}{2XQ})
\end{equation}
characterizes the gravitomagnetic field in this case. It is variable in magnitude but constant in direction (parallel to the $z$ axis). We note that $(2 X Q)^{-1}\ge 0$ vanishes along the axis of cylindrical symmetry and tends to $r_b/2$ with infinite slope at the outer boundary of the region under consideration here (see Fig.~\ref{fig:4}). The situation is more complicated, however, when we consider the gravitomagnetic components of the curvature tensor as measured by the fundamental observers. In particular, as demonstrated in Appendix~\ref{appen:A}, these do not all vanish along the symmetry axis. 

To illustrate these results further, let us consider a unit spacelike vector field $V^\mu$ that is carried by the fundamental observers and is  orthogonal to the $z$ axis; that is, 
\begin{equation}\label{eq:48n}
V^\mu=\cos\varphi\, \lambda^\mu_{ \hspace{0.07in}(r)}+\sin\varphi \,\lambda^\mu_{ \hspace{0.07in}(\phi)}.
\end{equation}
If this is a gyro axis, then it is Fermi-Walker transported along $x^\mu(T)$, namely, 
\begin{equation}\label{eq:49n}
\frac{D V^\mu}{dT}=(\mathcal{A}_\nu V^\nu)\lambda^\mu_{ \hspace{0.07in}(t)}.
\end{equation}
It follows from a detailed calculation that Eq.~\eqref{eq:49n} is equivalent to the condition that $d\varphi/dT=-\Omega^{(z)}$. Thus the gyro rotates with proper frequency $d\varphi/dT$ about the $z$ axis with respect to the spatial frame of the fundamental observers. 

The electric and magnetic components of the Riemann curvature tensor, projected on the tetrad frame of the fundamental observers, are given in Appendix~\ref{appen:A}.

\section{Geodesics}

Spacetime geodesics are generally obtained from the condition that the spacetime interval along the geodesic path be an extremal, namely, $\delta \int\,ds=0$,  where $ds^2$ is given by Eq.~\eqref{eq:8}. The geodesic equation then takes the form
\begin{equation}\label{eq:vn43}
\frac{d^2 x^\mu}{ds^2}+\Gamma^\mu_{\alpha\beta}\frac{d x^\alpha}{ds}\frac{d x^\beta}{ds}=0.
\end{equation}
In this section we treat  timelike and null geodesics in turn.

\subsection{Timelike Geodesics}
The geodesics of the dimensionless spacetime metric~\eqref{eq:8} depend on the choice of solution $X$ of the differential Eq.~\eqref{eq:fode} or the  
equivalent first-order system
\begin{equation}\label{eq:fodefo}
\frac{dX}{dr}=Y,\qquad \frac{dY}{dr}=-\frac{Y}{r^2 X^2}.
\end{equation}

For an admissible $X$, we choose its positive branch $X(r)>0$;  moreover,  to obtain the corresponding geodesic equations with respect to proper time $\tau$, we define $U:={dr}/{d\tau}$ and note the reparameterization of  system~\eqref{eq:fodefo}:  
\begin{equation}\label{eq:fodept}
\frac{dr}{d\tau}=U,\quad \frac{dX}{d\tau}=UY,\quad \frac{dY}{d\tau}=-\frac{UY}{r^2 X^2}.
\end{equation}
The components of the four-velocity vector of the free test particle along Killing vector fields are constants of geodesic motion. Therefore,  due to cylindrical symmetry, there are two constants of the motion $C_z=g_{z\alpha}dx^\alpha/d\tau$, the specific momentum in the $z$ direction, and $C_\phi=g_{\phi\alpha} dx^\alpha/d\tau$, the specific angular momentum about the $z$ axis; that is, 
\begin{equation}\label{eq:noether}
\frac{dz}{d\tau}=C_z e^t,\qquad \frac{d\phi}{d\tau}- X \frac{dt}{d\tau}= C_\phi r e^{-t}.
\end{equation}
By inserting these relations into the metric and simplifying, we find that
\begin{equation}\label{eq:vn46}
X^2 \Big(\frac{dt}{d\tau}\Big )^2=\frac{1}{r^3}\Big(\frac{dr}{d\tau}\Big )^2+\frac{X}{X_r}(e^{-t}+C_z^2 + C_\phi^2r e^{-2 t}),
\end{equation}
or
\begin{equation}\label{eq:dtdtau}
\frac{dt}{d\tau}=\frac{1}{X} V,  
\end{equation}
where
\begin{equation}\label{eq:V}
V:=\big[ \frac{1}{r^3} U^2+\frac{X}{X_r}(e^{-t}+C_z^2 + C_\phi^2r e^{-2 t})\big]^{1/2}.
\end{equation}
Our choice of positive sign in Eq.~\eqref{eq:dtdtau} is in conformity with the notion that the temporal coordinate should monotonically increase with proper time along the world line of an observer.
The Christoffel symbols $\Gamma^r_{\mu\nu}$ that are needed for the radial geodesic equation---namely, the component of Eq.~\eqref{eq:vn43} for the variation of the radial coordinate along the geodesic path---are given by 
\begin{align}\label{eq:chris}
\begin{split}
\Gamma^r_{tt}&=\frac{1}{2} r(-1+Q^2\frac{X}{Y}), \\ 
\Gamma^r_{tr}&=\Gamma^r_{rt}=\frac{1}{2},\\ 
\Gamma^r_{t\phi}&=\Gamma^r_{\phi t}=\frac{r X Q}{2 Y},\\ 
\Gamma^r_{rr}&=-\frac{1+3 r X^2 +r^2 XY}{2 r^2 X^2},\\
\Gamma^r_{\phi\phi}&=\frac{r X }{2 Y}.
\end{split}
\end{align}
Thus a full set of differential equations for the timelike geodesics can be expressed as
\begin{equation}\label{eq:fge}
\begin{split}
\frac{dt}{d\tau} &=\frac{1}{X} V,\\
\frac{dr}{d\tau} &=U,\\
\frac{d\phi}{d\tau} &= V+C_\phi r e^{-t},\\
\frac{dz}{d\tau} &=C_z e^t,\\
\frac{dU}{d\tau} &=-\Gamma^r_{tt}\frac{V^2}{X^2}-\frac{VU}{X} -2\Gamma^r_{t\phi}\frac{V}{X}( V+C_\phi r e^{-t})- \Gamma^r_{rr}U^2- \Gamma^r_{\phi\phi}( V+C_\phi r e^{-t})^2 ,\\
\frac{dX}{d\tau} &=UY,\\
\frac{dY}{d\tau} &=-\frac{UY}{r^2 X^2}.
\end{split}
\end{equation}

In a constant magnetic field configuration, the motion of a test charge is a combination of uniform rectilinear motion along the field direction together with uniform circular motion around this direction. The analogous situation in the gravitational field under consideration is, however, much more complex. In particular, the free motion of a test particle purely parallel to the rotation axis is impossible in the physical region $(r_b,\infty)$, since system~\eqref{eq:fge} does not have a solution for constant $r$ and $\phi$ coordinates. However, for a subset of admissible field configurations, special circular and helical motions are possible.  

In analogy with the magnetic case, let us look for geodesic motion that is confined to a cylinder of fixed radius $r>0$; that is, we let $U=0$ and $dU/d\tau=0$ in system~\eqref{eq:fge}.  The latter equation, after division by $(d\phi/d\tau)^2>0$, can be written as 
\begin{equation}\label{eq:60n}
\Gamma_{tt}^r \big(\frac{dt}{d\phi}\big)^2+ 2 \Gamma_{t\phi}^r \big(\frac{dt}{d\phi}\big)+\Gamma_{\phi\phi}^r=0,
\end{equation}
which is reminiscent  of the quadratic equation usually encountered in discussions of the gravitomagnetic clock effect~\cite{17}.  It is simple to show from Eqs.~\eqref{eq:chris} that
\begin{equation}\label{eq:61n}
(\Gamma_{t\phi}^r)^2-  \Gamma_{tt}^r\Gamma_{\phi\phi}^r=\frac{r^2 X}{4 Y}
\end{equation}
and
\begin{equation}\label{eq:62n}
\Gamma_{tt}^r= \frac{r X}{2 Y}\big [ Q-\big(\frac{Y}{X}\big)^{1/2}\big]\big[ Q+\big(\frac{Y}{X}\big)^{1/2}  \big].
\end{equation}
It follows from Eqs.~\eqref{eq:60n}--\eqref{eq:62n} that 
\begin{equation}\label{eq:63n}
\frac{dt}{d\phi}=-\frac{1}{Q \pm (Y/X)^{1/2}}.
\end{equation}
For an admissible solution $X$ of Eq.~\eqref{eq:fode}, $X(r)>0$, $Y(r)>0$ and $ Q(r)>0$; hence, the upper sign in Eq.~\eqref{eq:63n} would always lead to helical motion in the \emph{negative} sense about the $z$ axis, while for the lower sign the sense of the motion depends on the sign of 
\begin{equation}\label{eq:64n}
W(r):=  Q- \Big(\frac{Y}{X}\Big)^{1/2}.
\end{equation}
This function has been discussed in Sec. III; it is given by $A/(2 r_b^{1/2})$  at $r=r_b$ and asymptotically approaches $-a{S} _0>0$ as $r\to \infty$. In the class of admissible solutions, either $W(r)>0$ or $W$ has a zero $r_w$ in the physical interval $(r_b,\infty)$. This situation is illustrated in Fig.~\ref{fig:2}. In the former case, the sense of helical motion would always be negative,  while in the latter case we find from Eq.~\eqref{eq:60n}
\begin{equation}\label{eq:65n}
\big(\frac{dt}{d\phi}\big)_{r=r_w}=-\frac{1}{2 Q}.
\end{equation}
For $r\in ( r_b,r_w)$, however,  there could be helical motion in the \emph{positive} sense. This is crucial since we find from the relations for $dt/d\tau$ and $d\phi/d\tau$ in system~\eqref{eq:fge} that 
\begin{equation}\label{eq:66n}
\frac{dt}{d\phi}=\frac{V}{X ( V+C_\phi r e^{-t})}.
\end{equation}
In this relation we must have $C_\phi=0$,  since it follows from Eq.~\eqref{eq:63n} that $dt/d\phi $ must be constant for fixed $r$. Thus, $dt/d\phi =X^{-1}>0$ and from Eq.~\eqref{eq:63n}  we find that 
\begin{equation}\label{eq:67n}
-\frac{1}{ Q-(Y/X)^{1/2}}= \frac{1}{X}
\end{equation} 
holds for $rY= (Y/X)^{1/2}$, or 
\begin{equation}\label{eq:68n}
r^2 X Y=1.
\end{equation}
If this condition is satisfied for some $r_J \in(r_b,r_w)$,  then there is in general helical geodesic motion in the \emph{positive} sense about the $z$ axis regardless of the value of  $C_z \ne 0$. From the definition of  $F(r)$         in Eq.~\eqref{eq:F(r)}, we see that if           $F(r)=0$        has a solution  $r_J$  in the interior of the physical interval   $(r_b,\infty)$, then special helical solutions of the geodesic equation~\eqref{eq:vn43} exist; these special solutions reside in the \emph{gravitomagnetic jet}.

For $C_z=C_\phi=0$,  there is a special family of timelike circular geodesic orbits at $z=z_0$ with radius $r_J$ given by Eq.~\eqref{eq:68n},
\begin{equation}\label{eq:69n}
\phi=\phi_0+X(r_J)(t-t_0)
\end{equation} 
and
\begin{equation}\label{eq:70n}
e^{t/2}= e^{t_0/2}+\frac{1}{2} r_J (\tau-\tau_0),
\end{equation} 
where $t_0$, $\tau_0$, $\phi_0$ and $z_0$ are initial values of the corresponding quantities.
These circular orbits separate the up and down helical motions about the $z$ axis. Indeed, the special class of timelike geodesic orbits that define a gravitomagnetic jet include the special circular geodesics as limiting cases, for $C_z \to  0$, of helical motions parallel ($C_z > 0$) and
antiparallel ($C_z < 0$) to the axis of rotation. However, unless specified otherwise, such as in Sec.~VI, for example, the circular geodesic orbits of constant speed constitute a relatively negligible set of measure zero and are therefore generally ignored in our discussion of gravitomagnetic jets.

The existence of these special solutions of the timelike geodesic equation depends on whether there is a radial coordinate $r_J \in (r_b, \infty)$ for which $r_J^2 X(r_J) X_r(r_J)=1$. Given an admissible solution $X$---two examples of which are depicted in Fig.~\ref{fig:2}---one can show that either there is only one such $r_J$ or there is none. To see this, we simply note that the function $(r^2 X_r)^{-1}>0$ starts from zero  at $r=r_b$ with infinite slope, has a maximum at $2 r X^2=1$ and then drops off to zero as $r\to \infty$. Thus, $X$ and $(r^2 X_r)^{-1}$ either do not intersect each other for  $r\in (r_b, \infty)$ or they do so at exactly one point. 

By imposing the conditions  that  $C_\phi=0$ and $r=r_J$, it follows immediately that $r$, $U$, $X$ and $Y$ remain at their initial values---in particular, the right-hand side of the differential equation for $dU/d\tau$ vanishes---while 
\begin{align}\label{eq:rodeg}
\begin{split}
\frac{dt}{d\tau} &=r_J  (e^{-t}+C_z^2)^{1/2},\\
\frac{d\phi}{d\tau}&=r_J X(r_J ) (e^{-t}+C_z^2)^{1/2},\\
\frac{dz}{d\tau}&=C_z e^t.
\end{split}    
\end{align} 
This system can be reduced to quadrature in elementary functions. For $C_z\ne 0$,  its flow is given by
\begin{align}\label{eq:flowg}
\begin{split} t&= 2\ln(\frac{1}{|C_z|}\sinh [\frac{1}{2} r_J |C_z|(\tau-\tau_0)+\sinh^{-1}({|C_z|}e^{t_0/2})]),\\
\phi&=\phi_0+X(r_J)(t-t_0) ,\\
z&=z_0+\frac{1}{C_z}\int_{\tau_0}^\tau \sinh^2 [\frac{1}{2} r_J |C_z|(\tau'-\tau_0)+\sinh^{-1}({|C_z|}e^{t_0/2})] \, d\tau',
\end{split}    
\end{align}
where we consistently use zero subscripts to denote initial values. It is clear that $t$ and $\phi$ increase as $\tau-\tau_0$ increases; however, $z-z_0$ increases for $C_z>0$ and decreases for $C_z<0$, thereby leading to gravitomagnetic jets propagating up and down parallel to the rotation axis. The function $z(t)$ may be expressed as 
\begin{equation}\label{eq:vn63} z(t)=z_0+K^+(t)-K^+(t_0),\end{equation}
where
\begin{align}\label{eq:vn64}
\begin{split}
K^+&=\frac{C_z}{r_J|C_z|^3}
\{ \kappa^+(1+{\kappa^+}^2)^{1/2}-\ln [{\kappa^+}+(1+{\kappa^+}^2)^{1/2}]\},\\
\kappa^+&=|C_z|e^{t/2}.
\end{split}
\end{align}

\subsection{Null Geodesics}
Let $\zeta$ be an affine parameter along the world line of a null geodesic. It follows from the existence of the spacelike Killing vectors $\partial_z$ and $\partial_\phi$ that 
\begin{equation}\label{ng1}
\frac{dz}{d\zeta}=\hat C_z e^t, \quad \frac{d\phi}{d\zeta}-X \frac{dt}{d\zeta}=\hat C_\phi r e^{-t},
\end{equation}
where $\hat C_z$ and $\hat C_\phi$ are constants of the motion. The spacetime path is null ( $ds^2=0$); hence, 
\begin{equation}\label{ng2}
X^2\big( \frac{dt}{d\zeta}\big )^2= \frac{1}{r^3} \big( \frac{dr}{d\zeta}\big )^2 + \frac{X}{X_r} (\hat C_z^2+\hat C_\phi^2 r e^{-2t}).
\end{equation}
Let us define $\hat U$ and $\hat V$ such that $\hat U=dr/d\zeta$ and
\begin{equation}\label{ng3}
\hat V:= \big[  \frac{1}{r^3} \hat U^2+ \frac{X}{X_r} (\hat C_z^2+\hat{C}_\phi^2 r e^{-2t}) \big]^{1/2}.
\end{equation}
Then, the geodesic equations for a null path are
\begin{equation}\label{ng4}
\begin{split}
\frac{d t}{d\zeta}={}& \frac{\hat V}{X}, \\
\frac{d r}{d\zeta}= {}&{\hat U}, \\
\frac{d \phi}{d\zeta} ={}&{\hat V} +\hat C_\phi r e^{-t}, \\
\frac{d z}{d\zeta}={}&\hat C_z e^t, \\
\frac{d\hat U}{d\zeta}={}& -\Gamma_{tt}^r \frac{\hat V^2}{ X^2}- \frac{\hat U \hat V}{X} - 2 \Gamma_{t\phi}^r \frac{\hat V}{X} (\hat V+\hat C_\phi r e^{-t}) \\
& - \Gamma_{rr}^r \hat U^2-\Gamma_{\phi\phi}^r (\hat V+\hat C_\phi r e^{-t})^2,\\
\frac{dX}{d\zeta}={}&\hat U Y,\\
\frac{dY}{d\zeta}={}&-\frac{\hat U Y}{r^2 X^2}.
\end{split}
\end{equation}

As before, with $\hat C_\phi=0$, it is possible to find an exact class of solutions of these equations once $r^2 X X_r=1$ for some $r_J\in (r_b,\infty)$. Each null geodesic in this special class follows a helical trajectory in the positive sense on the cylindrical surface $r=r_J$ with $\hat C_z\ne 0$ and 
\begin{equation}\label{ng5}
\begin{split}
t-t_0 &= r_J | \hat C_z| (\zeta-\zeta_0), \\
\phi-\phi_0 &= r_JX( r_J)| \hat C_z| (\zeta-\zeta_0),\\
z-z_0 &= \frac{1}{  r_J }\frac{ \hat C_z}{ | \hat C_z| }(e^t-e^{t_0} ).
\end{split}
\end{equation}
For the example of $X$ depicted in the middle panel of Fig.~\ref{fig:2},  $ r_J\approx 0.7739$,  $X( r_J )\approx 0.4288$ and  $X_r( r_J )\approx 3.8941$. Let us observe that $z(t)$        for the special null geodesics coincides with the late-time behavior of special timelike geodesics with $C_z\ne 0$, since   $K^+$    is given, as   $t\to \infty$, by
\begin{equation}\label{eq:vn68}
K^+\sim \frac{C_z}{r_J|C_z|}e^t.
\end{equation}

We remark that the special case of null circular geodesics is excluded here; that is, $\hat C_z=\hat C_\phi=0$ is not possible, since $dt/d\zeta$ would then vanish and this is forbidden.

A characteristic feature of the special (timelike and null) solutions of the geodesic equations is that the sense of helical motion is always positive. This is due to our choice of the positive branch of solutions of Eq.~\eqref{eq:fode}, namely, $X>0$.  In fact, metric~\eqref{eq:8} remains invariant under $X\mapsto -X$ and $\phi\mapsto -\phi$. Thus if we work exclusively with the negative branch   $X<0$   instead, the sense of helical motion will be negative. Hence the significant feature of gravitomagnetic jets that must be emphasized here is simply that in the double-jet configuration, both jets have the \emph{same} helical sense. Moreover, it follows from the solutions of the equations of jet motion that there is a characteristic \emph{exponential} dependence of       
 $|z-z_0|$  upon the azimuthal angle $\phi-\phi_0$ in gravitomagnetic jets. It is important to emphasize that the radius of the helical path of a gravitomagnetic jet is constant only in terms of $r$; in fact, the helix expands as its proper radius is given by $\exp(t/2)R_J$. We note that helical motions in astrophysical jets have been the subject of recent investigations---see, for instance, \cite{nn22, nn23} and references therein.

\subsection{Jets}
It is important to note that for $t\to \infty$, the special helical timelike geodesics  approach  the special null geodesics asymptotically; in fact, this can be simply seen from the formal correspondence between the respective geodesic equations.  That is, systems~\eqref{eq:fge} and~\eqref{ng4}  become \emph{formally} equivalent  once $V$ and $\hat V$ in Eqs.~\eqref{eq:V} and~\eqref{ng3}, respectively, take the same form; this actually happens when $\exp(-t)\to 0$ in Eq.~\eqref{eq:V}. To gain physical perspective, however, let $u^\mu=dx^\mu/d\tau$ be the four-velocity vector of a free test particle following a \emph{special} timelike geodesic. With regard to the fundamental observers along the path of the particle, 
\begin{equation}\label{eq:84}
 u^\mu=u^{(\alpha)}\lambda^\mu_{ \hspace{0.07in}(\alpha)},
\end{equation}
where $u^{(\alpha)}=\gamma(1,\mathbf{v})$, $\mathbf{v}$ is the local velocity of the particle as measured by the fundamental observers and $\gamma$ is the corresponding Lorentz factor. Thus $\gamma=-u_\mu\lambda^\mu_{ \hspace{0.07in}(t)}$, which can be calculated using Eq.~\eqref{eq:41n} and the result is
\begin{equation}\label{eq:85}
\gamma=\Big(\frac{1+C_z^2 e^t}{1+H(r_J)}\Big)^{1/2},
\end{equation}
where $H$ is given by Eq.~\eqref{eq:21m}.

For a jet,   $C_z\ne 0$            and hence the Lorentz factor for a jet is always larger than            $\gamma_{\mathtt{min}}$, which is defined by Eq.~\eqref{eq:85} with $C_z=0$. Thus according to the fundamental observers,        $\gamma_{\mathtt{min}}$ is the Lorentz factor for a free test particle on a circular orbit of radius  $r_J$    about the axis of cylindrical symmetry. Let us recall that for admissible solutions of Eq.~\eqref{eq:fode},    $rXQ>0$                and hence   $1+H(r)>0$                     for all              $r\in (r_b,\infty)$. When jets are allowed in an admissible solution, there exists a unique  $r_J\in (r_b,\infty)$                    for which       $F(r_J)=0$. From   $H=F-rX^2$, we find that                                   $-1<H(r_J)=-r_JX^2(r_J)$, so that $\gamma_{\mathtt{min}}$         corresponds to a scale-invariant minimum speed  $\beta_{\mathtt{min}}$            given by
\begin{equation}\label{eq:100n}
\beta_{\mathtt{min}}^2=r_JX^2(r_J).
\end{equation}
Thus the jet speed is always greater than $ \beta_{\mathtt{min}}$. For the solution displayed in the middle panel  of Figure~\ref{fig:2}, $ \beta_{\mathtt{min}}\approx 0.3772$. In general, the local velocity of the particle as determined by the fundamental observers can be written as $\mathbf{v}=(v_r,v_\phi,v_z)$, where $v_r=0$, $v_\phi=\beta_{\mathtt{min}}$ and $\gamma v_z=C_z \exp (t/2)$.  It is clear from Eq.~\eqref{eq:85} that for $C_z\ne 0$, $\gamma$ diverges as $t\to \infty$, so that the local jet speed asymptotically approaches the speed of light according to the fundamental observers. This circumstance  comes about due to the specific exponential dependence of metric~\eqref{eq:8} upon time $t$. This fascinating dynamical feature of the gravitational field is presumably caused by an exterior configuration whose characterization would necessitate a separate investigation. As a free test particle in a jet follows a helical path either up or down on a cylinder of radius $r_J$, the gravitational potentials along its world line vary in time in just such a way that the particle is apparently accelerated with its speed approaching the light speed for $t\to \infty$. For such extremely energetic test particles, however, our approximation scheme may break down at some point; that is, the gravitational influence of the test particle on the background spacetime may no longer be negligible.

The preceding considerations may be employed to illustrate certain asymptotic characteristics of the gravitomagnetic jets using Eq.~\eqref{ng5}. Let $z^*$ be the proper distance parallel to the axis of cylindrical symmetry; then,  $dz^*=\exp(-t/2)\, dz$ from Eq.~\eqref{eq:8}. Moreover, an appropriate radial coordinate is $\rho^*=\rho \exp(t/2)$. Therefore, along the special null geodesics we have
\begin{equation}
\label{eq:87} \frac{dz^*}{dt}=\frac{\hat{C}_z}{r_J|\hat{C}_z|}e^{t/2},\qquad
\rho^*=r_J^{-1/2}\, e^{t/2}.
\end{equation}
It follows that 
\begin{equation}\label{eq:90}
\Big|\frac{\rho^*-\rho_0^*}{z^*-z_0^*}\Big|=\frac{1}{2} r_J^{1/2}.
\end{equation}

How can one numerically find admissible solutions of Eq.~\eqref{eq:fode} that allow jets?  For every  solution with or without a jet, scaling  provides a one-parameter family of solutions of the same kind.   That is, every jet solution belongs to a one-parameter family of solutions due to the scaling property of Eq.~\eqref{eq:fode}; indeed, this differential equation remains invariant under the transformation $r\mapsto \sigma r$ and $X(r)\mapsto \sigma^{-1/2} X(r)$ for $\sigma\in (0,\infty)$. The scaled solution is defined over the interval $(\sigma r_b,\infty)$; moreover,
$Q\mapsto Q /\sigma^{1/2}$, $W\mapsto W/\sigma^{1/2}$, $H\mapsto H$ and $F\mapsto F$. In particular, if a jet exists in the original solution at $r_J$, the new scaled solution has a jet at $\sigma r_J$. This property may thus be employed to set $r_J=1$ for every jet solution. One can then numerically integrate  Eq.~\eqref{eq:fode} with initial conditions such that $r_J=1$, $X(1)=1/\vartheta$ and $X_r(1)=\vartheta$, where $\vartheta\ge \vartheta_{\mathtt{min}}$. Here $\vartheta_ {\mathtt{min}}^{-1}$ is the maximum allowed value of $\beta_{\mathtt{min}}$, which according to Fig.~\ref{fig:w1} of Appendix~\ref{appen:B} is about $0.63$; hence, $\vartheta_ {\mathtt{min}}\approx 1.6$.  All such solutions are admissible according to the arguments presented in Appendix~\ref{appen:B}. Moreover, $r_b\approx 1-\vartheta^{-2}/2$ for $\vartheta\gg 1$ in accordance with Eq.~\eqref{eq:bm26}. As $\vartheta \to \infty$, all (azimuthal) helical motions disappear and the special timelike and null geodesics become vertical.

Finally, along the special timelike geodesic path with $u^\mu=dx^\mu/d\tau$ given by Eq.~\eqref{eq:rodeg}, consider an observer that carries  an orthonormal parallel-propagated tetrad frame $\Lambda^\mu_{ \hspace{0.07in}(\alpha)}$ with $\Lambda^\mu_{ \hspace{0.07in}(0)}=u^\mu$ and a spatial frame given by a set of unit gyro axes that can be expressed in $(t,r,\phi,z)$ coordinates as 
\begin{align}
\label{xxx71} \Lambda^\mu_{ \hspace{0.07in}(1)}=&(0,\;r_J^{5/2}X(r_J)e^{-t/2}\cos\frac{t}{2}, \;r_J^{1/2}e^{-t/2} \sin\frac{t}{2},\;0),\\
\label{xxx72} \Lambda^\mu_{ \hspace{0.07in}(2)}=&(0,\;r_J^{5/2}X(r_J)e^{-t/2}\sin\frac{t}{2}, \;-r_J^{1/2}e^{-t/2} \cos\frac{t}{2},\;0),\\
\label{xxx73} \Lambda^\mu_{ \hspace{0.07in}(3)}=&(r_J C_z,\;0, \;r_J X(r_J)C_z,\; e^t(e^{-t}+C_z^2)^{1/2}).
\end{align}
A free pointlike test gyroscope with spin $S^\mu$ carried by the observer along the path is then given by $S^\mu=S^{(i)} \Lambda^\mu_{ \hspace{0.07in}(i)}$, where $S^{(i)}$, $i=1,2,3$, are constants. It is straightforward to verify that the requirements of orthogonality ($u_\mu S^\mu=0$) and parallel transport ($DS^\mu/d\tau=0$) are satisfied; moreover, $S_\mu S^\mu=S_{(i)}S^{(i)}$ is a constant of the motion. The spin vector in general undergoes damped precessional motion of frequency $\frac{1}{2}$ with respect to time $t$; by contrast, the orbital frequency of geodesic motion is $X(r_J)$. The precessional motion decays exponentially; in fact, as $t\to \infty$ the special timelike geodesic  with $C_z\ne0$ approaches a null geodesic and $S^\mu\sim (S^{(3)} C_z/|C_z|) u^\mu$, as expected~\cite{23}. This follows from the fact that for $t\to \infty$, $ \Lambda^\mu_{ \hspace{0.07in}(1)}$ and $ \Lambda^\mu_{ \hspace{0.07in}(2)}$ asymptotically  tend to zero and  $ \Lambda^\mu_{ \hspace{0.07in}(3)}\sim (C_z/|C_z|) u^\mu$. The projection of the curvature tensor on $  \Lambda^\mu_{ \hspace{0.07in}(\alpha)}$ turns out to involve somewhat complicated functions of time $t$. In principle, the curvature components as measured by the observer may be used to study the generalized Jacobi equation~\cite{4} along the special geodesic world line, but that is beyond the scope of this paper.

\section{Numerical Results: Jets are Attractors}
\begin{figure}
\centerline{\psfig{file=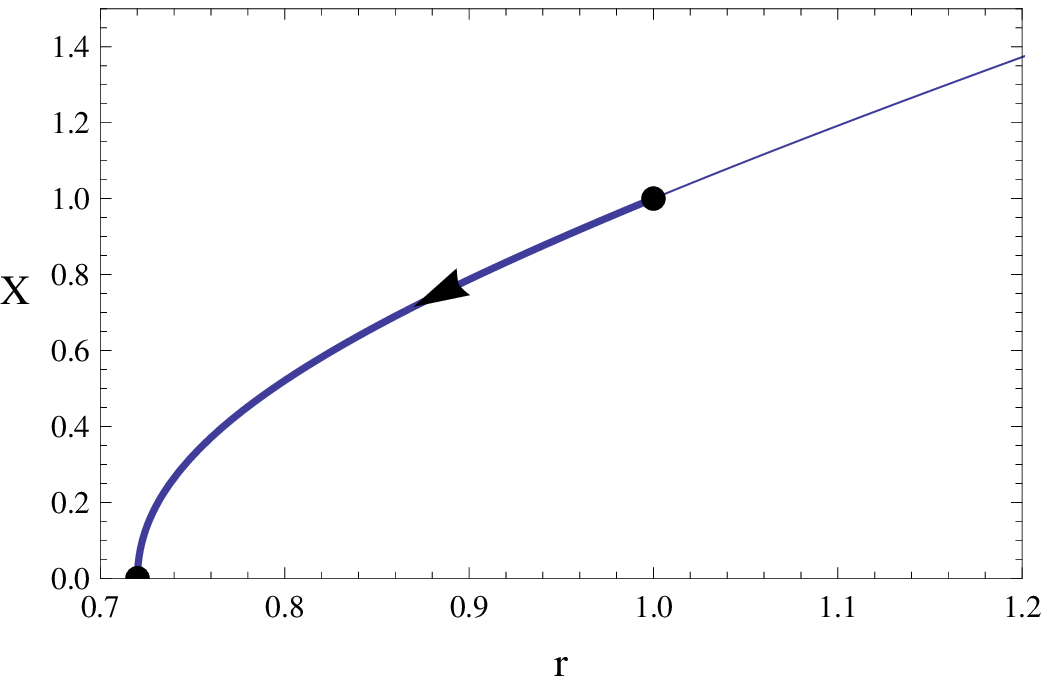, width=15pc}\qquad \psfig{file=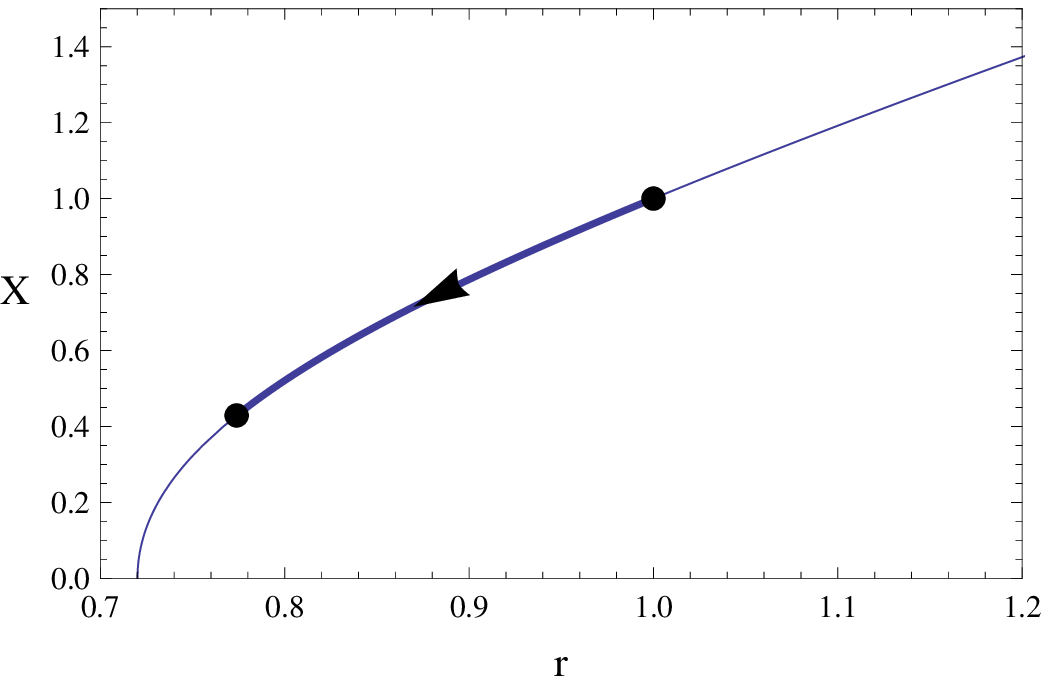, width=15pc}}
\caption{\label{fig:rho} Schematic diagram  illustrating the results of our numerical work. Right panel: For a range of parameters close to the jet parameters, the radial coordinate \emph{as a function of the particle's proper time} starts from its initial value $r_0$ and reaches $r_J$. It is then fixed at $r_J$ while the test particle executes helical motion up or down parallel to the rotation axis as in jets (see Fig.~\ref{fig:spiral}). Left panel: Beyond a certain range in the canonical angular momentum $C_\phi$, the test particle cannot be confined; that is, $r_J$ is bypassed and the particle soon reaches the boundary cylinder, thus leaving the spacetime region of interest. In constructing this figure, we have used the
solution $X$ given in the middle panel of Fig.~\ref{fig:2}; therefore, $r_b \approx 0.720$, $r_J
\approx 0.774$ and $r_0 = 1$.
 }
\end{figure}

\begin{figure}
\centerline{\psfig{file=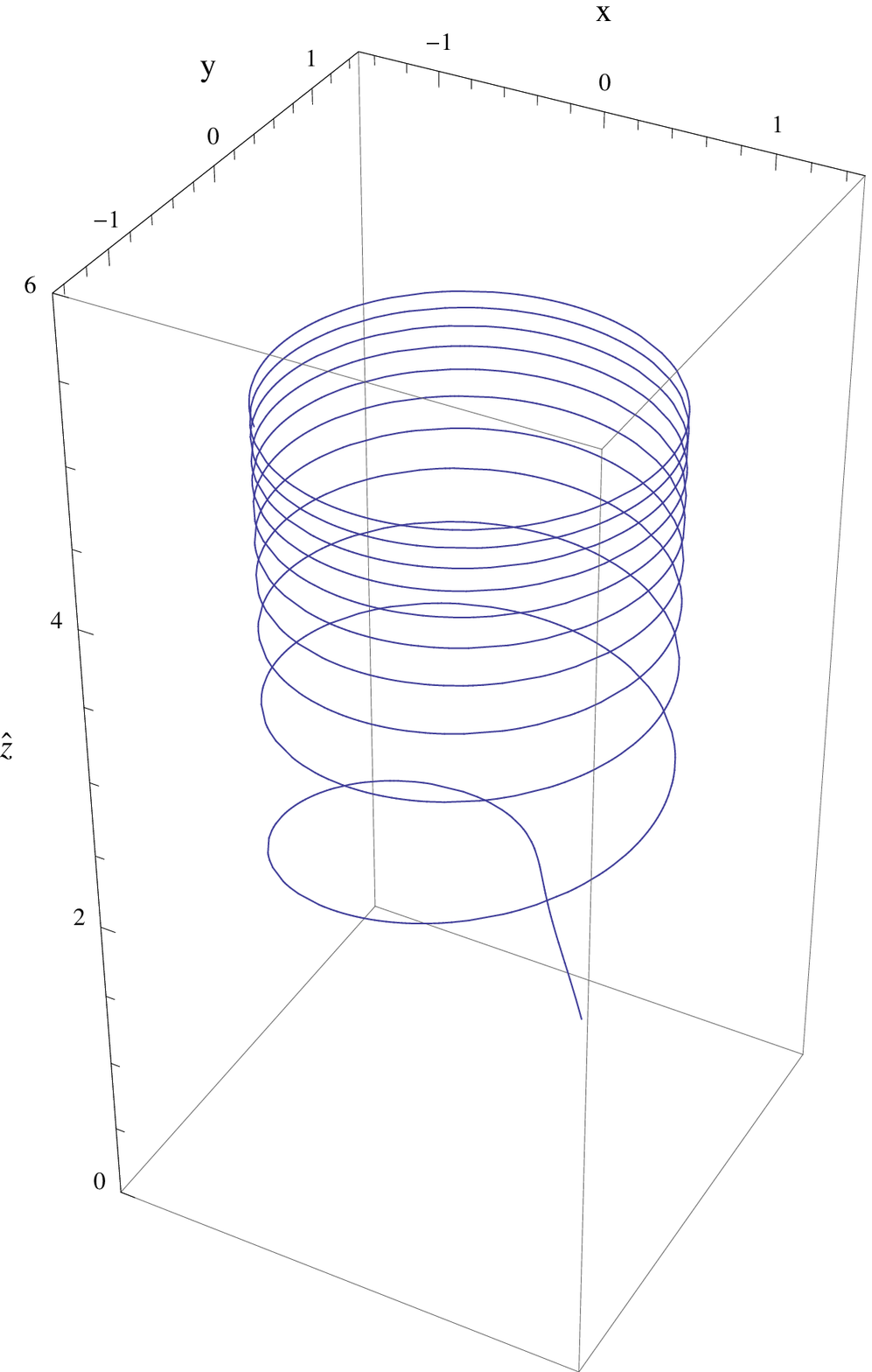,width=20pc} \psfig{file=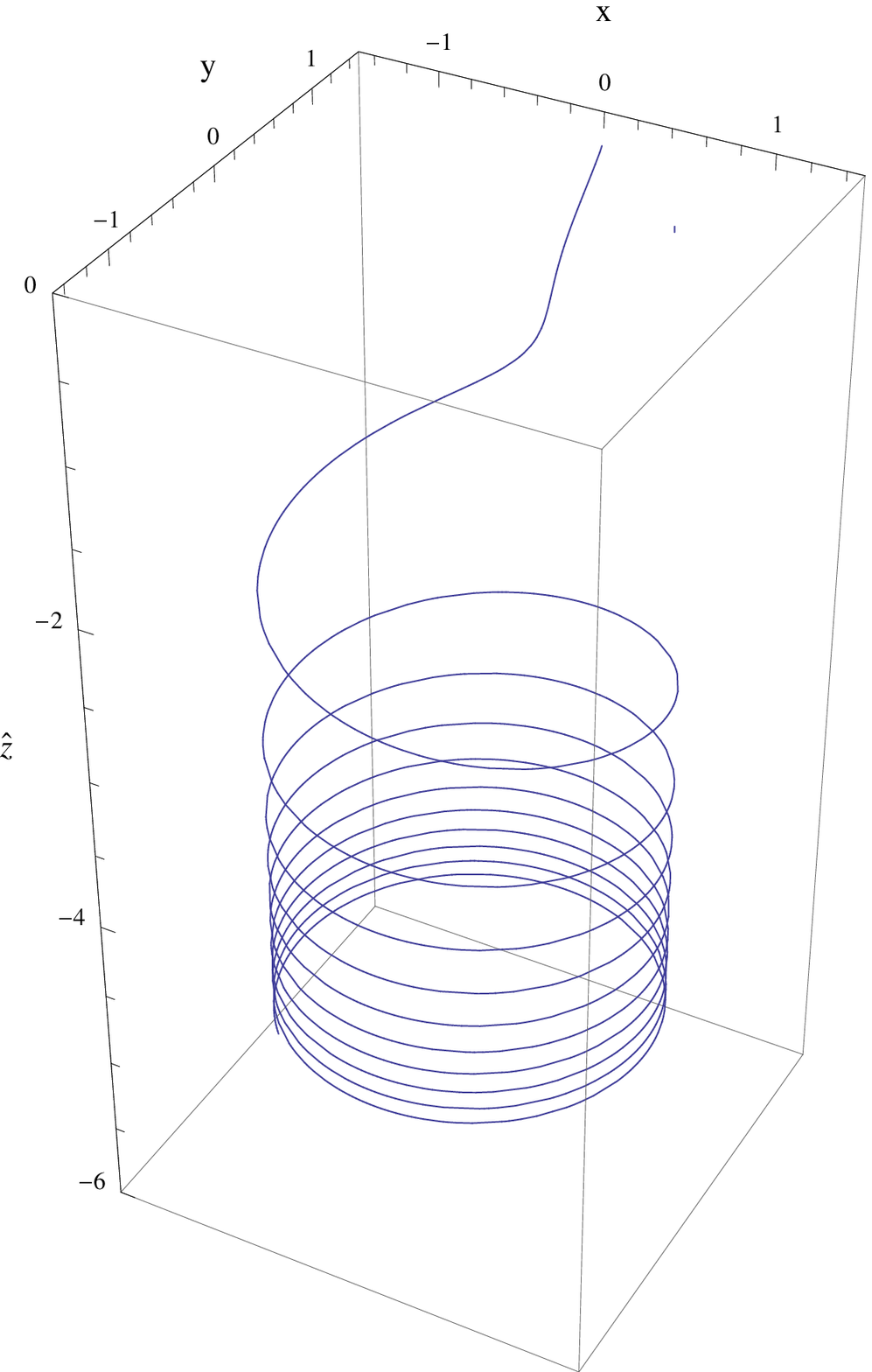,width=20pc}}
\caption{\label{fig:spiral} The result of integration of system~\eqref{eq:fge} for timelike geodesics attracted to jets. The initial data at $\tau_0=0$ are  
$t_0 = 0$,
$r_0 = 1$, $U_0 = 0$,
$\phi_0 = 0$,
$z_0 = 0$,
$X_0 = 1$ and 
$Y_0 = 2$. The parameters are $C_\phi = 0.9$ and
$C_z = \pm 1$. The left-hand plot (jet going up) is for $C_z=1$ and the right-hand plot (jet going down) is for $C_z=-1$. The coordinates are $(x,y,\hat z)$, where $x=\rho\cos\phi$, $y=\rho\sin\phi$ and $\rho=r^{-1/2}$; moreover, $\hat z=\ln|\ln z|$ for the left-hand plot and $\hat z=-\ln|\ln |z||$ for the right-hand plot. We use $\hat z$ instead of $z$ for the sake of clarity.
}
\end{figure}

The special exact solutions of the geodesic equations that correspond to jets have vanishing (canonical) angular momentum $C_\phi=0$ and are confined to cylindrical surfaces of fixed $r=r_J$; otherwise, they have arbitrary $C_z$, $t_0$, $\phi_0$ and $z_0$ for an admissible solution $X$ that allows jets; for example, the one with $X(1)=1$ and $X_r(1)=2$ given in the middle panel of Fig.~\ref{fig:2}. For this $X$, we have numerically integrated system~\eqref{eq:fge} with $C_z=\pm 1$ and initial data $t_0=0$, $r_0=1$, $U_0=0$, $\phi_0=0$ and $z_0=0$ at $\tau_0=0$; we find that from $C_\phi=0$ to $C_\phi=0.9$, solutions are attracted to the jet while for $C_\phi\ge 1$,  solutions are not attracted to the jet. The numerical results are presented in Figs.~\ref{fig:rho} and~\ref{fig:spiral}. 

More generally, our numerical experiments suggest that the codimension-two submanifold 
\begin{equation}\label{eq:Ng}
\{\text{$(t,r,\phi,z,U)$: $r=r_J$ and $ U=0$}\}    
\end{equation}
is a gravitomagnetic jet.  The flow on the invariant manifold~\eqref{eq:Ng} has a simple geometric interpretation in physical space: each geodesic remains at a fixed radius $r_J$ from the axis of cylindrical symmetry. Solutions with  $C_z\ne 0$ spiral around the $z$ axis on a helix that becomes unbounded as proper time approaches infinity.  The measure-zero set of solutions with $C_z=0$ remains bounded in circular motion about the $z$ axis.  While we have not explored the entire parameter and state spaces,  our numerical experiments confirm that the manifold~\eqref{eq:Ng} attracts all nearby geodesics. Thus,  our experiments suggest that after transient motions nearby geodesics spiral about the $z$ axis with radii approaching $r_J$ and, except for the negligible set with $C_z=0$,  become unbounded as proper time approaches infinity.

More precisely, in our four-dimensional spacetime the geodesic flow takes place in  an eight-dimensional state space (the tangent bundle of the spacetime).
By the introduction of proper time, we consider only unit-speed 
geodesics. This reduces the state space to seven dimensions. Cylindrical symmetry implies  that there are 
two integrals of the motion $C_\phi$ and $C_z$; the state space is thereby reduced to five dimensions $(t, r, \phi, z, U)$. Since the jet is confined to the cylinder $r=r_J$, it is a three-dimensional manifold parameterized by $(t, \phi, z)$.  We also note that the third and fourth differential equations for $\phi$ and $z$ in the geodesic equations~\eqref{eq:fge} may be decoupled from this system. Thus, the attraction properties of the manifold of special solutions correspond to the behavior of $r$ and $U$. A geodesic is attracted to the manifold~\eqref{eq:Ng} if $r$ approaches $r_J$ and $U$ approaches zero as $\tau\to \infty$.   

\section{Discussion}
The cylindrical symmetry employed throughout this work has been a useful simplifying assumption. However, axial symmetry is expected to be a better approximation for the treatment of high-energy astrophysical jets that appear in circumstances involving a rotating collapsed configuration surrounded by a rotating accretion disk. On the other hand, in this case an analytical treatment appears to be prohibitively complicated.

How could one generate the gravitational fields discussed in this paper? In electrodynamics, the magnetic field $\mathcal{B}$ inside an infinite circular cylinder is uniform and parallel to the axis of symmetry, provided the cylinder is surrounded by a uniform current sheet. The connection between the source and the interior field is given by $ \mathcal{B} = 4 \pi i /c$, where $i$ is the amount of electric current per unit length of the cylinder. This is a good approximation around the axis near the center of a long solenoid. In our gravitational case, we would expect that the source-free dynamic interior solution could be joined---perhaps along the inner and outer boundaries---to an exterior solution that could serve as the source of the interior field.  However, finding such a source requires a separate investigation that is beyond the scope of this paper. In fact, it may be advantageous to look for gravitomagnetic jets in more realistic axisymmetric systems in general relativity.

Despite these drawbacks, it is remarkable that general relativity permits the existence of Ricci-flat rotating cylindrical gravitational fields that admit gravitomagnetic jets, which are formed from solutions of the geodesic equation corresponding to generally helical motions of free test particles up and down parallel to the axis of symmetry with speeds that asymptotically approach the speed of light.  Particle acceleration mechanisms are important in astrophysics~\cite{vn24}. We have shown that general relativity can in principle provide a purely gravitational mechanism for the directional acceleration of test particles to ultrarelativistic speeds.

\appendix
\setcounter{section}{0}
\newcounter{saveeqn}%
\newcommand{\alpheqn}{\setcounter{saveeqn}{\value{equation}}
\stepcounter{saveeqn}\setcounter{equation}{0}
\renewcommand{\theequation}{\mbox{\Alph{section}\arabic{equation}}}}
\renewcommand{\thesection}{\Alph{section}}
\section{Curvature Tensor}\label{appen:A}
\alpheqn
For the Ricci-flat spacetime metric represented by Eq.~\eqref{eq:8} and $x^\mu=(t,r,\phi,z)$, the metric tensor and its inverse are given by 
\begin{align}
\label{A1} 
(g_{\mu\nu})&=
\left[
\begin{array}{cccc}
-\frac{e^t XQ}{r} & 0 & -\frac{e^t X}{r} & 0\\
0 & \frac{e^t X_r}{r^3 X} & 0 & 0\\
-\frac{e^t X}{r} & 0 &\frac{e^t }{r} &0\\
0 & 0& 0& e^{-t }
\end{array}
\right],\\
\label{A2}
(g^{\mu\nu})&=
\left[
\begin{array}{cccc}
-\frac{e^{-t}}{XX_r} & 0 & -\frac{e^{-t}}{X_r} & 0\\
0 & \frac{e^{-t}r^3 X}{ X_r} & 0 & 0\\
-\frac{e^{-t}}{X_r} & 0 & \frac{e^{-t}Q }{ X_r} &0\\
0 & 0& 0& e^t 
\end{array}
\right].
\end{align}

The nonzero components of the connection, modulo its symmetry ($\Gamma_{\mu\nu}^\alpha=\Gamma_{\nu\mu}^\alpha$), can be expressed as
\begin{align}
\label{A3} \Gamma_{tt}^t&=\frac{1}{2} (1+ \frac{X}{r X_r}), \qquad \Gamma_{tr}^t=\frac{H}{2 r^2 X^2},\\
\label{A4} -\Gamma_{t\phi}^t&= X\Gamma_{\phi\phi}^t=\frac{X}{X_r}\Gamma_{r\phi}^t =- \frac{1}{X Q}\Gamma_{tt}^\phi=\frac{1}{Q}\Gamma_{t\phi}^\phi=\Gamma_{\phi\phi}^\phi=\frac{1}{2 r X_r},\\
\label{A5} X\Gamma_{rr}^t&=-\frac{1}{r}\Gamma_{tr}^\phi = \Gamma_{rr}^\phi=\frac{1}{2 r^3 X},\\
\label{A6} X\Gamma_{zz}^t&=\Gamma_{zz}^\phi=-\frac{e^{-2t}}{2X_r}, \qquad \Gamma_{tz}^z=-\frac{1}{2},
\end{align}
together with the $\Gamma_{\mu\nu}^r$ components given in Eq.~\eqref{eq:chris} of Sec. V. We note that $Q$     and  $H$    have been defined in Eq.~\eqref{eq:admiss} and Eq.~\eqref{eq:21m}, respectively.

The components of the curvature tensor projected on the tetrad frame of fundamental observers, 
\begin{equation}\label{A7}
R_{(\alpha)(\beta)(\gamma)(\delta)}=R_{\mu\nu\rho\sigma}\lambda^\mu_{ \hspace{0.07in}(\alpha)}\lambda^\nu_{ \hspace{0.07in}(\beta)}\lambda^\rho_{ \hspace{0.07in}(\gamma)}\lambda^\sigma_{ \hspace{0.07in}(\delta)},
\end{equation}
can be represented as a $6\times 6$ matrix in the standard manner,
\begin{equation}\label{A8}
\mathcal{R}=\left[
\begin{array}{cc}
\mathcal{E} &\mathcal{H}\\
\mathcal{H}& -\mathcal{E}
\end{array}
\right],
\end{equation}
where $\mathcal{E}$ and $\mathcal{H}$ are symmetric and traceless $3\times 3$ matrices in a Ricci-flat spacetime. We identify  $\mathcal{E}$ and $\mathcal{H}$  respectively with the electric and magnetic components of spacetime curvature according to the fundamental observers. The tidal matrix $\mathcal{E}$ is given by
\begin{equation}\label{A9}
\mathcal{E}=\frac{e^{-t}}{4Q X X_r}\left[
\begin{array}{ccc}
Q+3 X & -P & 0\\
-P& Q & 0\\
0 & 0  & -2Q-3 X
\end{array}
\right],
\end{equation}
where $P=(Q/(r X))^{1/2}$.

The magnetic part of the curvature is given by
\begin{equation}\label{A11}
\mathcal{H}=\frac{e^{-t}}{4Q }\big(\frac{r}{XX_r }\big)^{1/2}\left[
\begin{array}{ccc}
0 & 0 & 3\\
0& 0 & \mathcal{P}\\
3 & \mathcal{P} &0 
\end{array}
\right],
\end{equation}
where $\mathcal{P}$ can be expressed, using $H=r X Q-1$, as 
$
\mathcal{P}=HP /X
$.

Near the symmetry axis, $r\to \infty$, we have 
$X\sim ar$,  $X_r\sim a$ and $Q\sim -aS_0$,
so that $e^t \mathcal{E}$ and $e^t \mathcal{H}$ are in general nonzero constant matrices. 
Similarly, $e^t \mathcal{E}$ and  $e^t \mathcal{H}$ are in general nonzero constant matrices at the boundary cylinder as well. That is, near the boundary $r\to r_b$, $X\to 0$, $X_r\to\infty$ and $F\to 0$. Moreover, $H=F-r X^2$ and using Eq.~\eqref{eq:20m}, we find that 
\begin{equation}\label{eq:A14}
\Big(\frac{\mathcal{P}}{Q}\Big)_{r=r_b}=A r_b^{1/2}.
\end{equation} 
We recall that $A<0$ when jets are present and $A\ge 0$ in the absence of jets. In the special case of $A=0$, $e^t \mathcal{H}$ vanishes at the boundary; moreover, it is possible to show that for $r\to r_b$, 
\begin{equation}\label{eq:A15}
X=\pm \big(\frac{2\epsilon}{r_b}\big)^{1/2} (1+\frac{1}{2}\epsilon-\frac{5}{72}\epsilon^2+O(\epsilon^3)),
\end{equation}
where $r=r_b(1+\epsilon)$. This result corrects an error in Eq. (34) of~\cite{1}, where $3/76$ occurs in place of $5/72$.

\section{Properties of solutions of $r^2 X^2 X_{rr}+X_r=0$}\label{appen:B}

To avoid confusion, we emphasize that some of the notation employed in this appendix is specific to the mathematical arguments at hand and is independent of the rest of the appendixes or this paper.

We begin with two obvious facts about the solutions of the differential Eq.~\eqref{eq:fode}: If $X$ is a solution, then $-X$ is  a solution. Also, we have that 
\begin{equation}\label{eq:factB}
\frac{d}{dr}(X_r^2)=-2\Big(\frac{X_r}{ r X}\Big)^2.
\end{equation}

Given $r_0>0$ and initial data $X(r_0)>0$ and $X_r(r_0)>0$, there is, by the usual existence theory,  a unique solution of the differential Eq.~\eqref{eq:fode} defined on some maximal interval $(r_b,r_B)$ with $r_b<r_0<r_B$. In fact, $r_B=\infty$. To prove this, we may use the extension theorem for ordinary differential equations (see, for example,~\cite{ccc}). In effect, a solution continues to exist until it reaches the domain of definition of the differential equation or it blows up to infinity. By Eq.~\eqref{eq:factB} and for our choice of (positive) initial data,  $X_r$ is a monotonically decreasing function of $r$ as long as $X(r)$ and $X_r(r)$ are both positive. If either of these functions vanishes for some $r>r_0$, then there must be a point where $X_r$ vanishes. Let $r_\omega$ be the infimum of all such points and note  that by continuity $X_r(r_\omega)=0$. The function $X(r)\equiv X(r_\omega)$ is clearly a solution of the differential equation that has the same (initial) data at $r_\omega $. By uniqueness the two solutions must be the same. But, this is a contradiction: the original solution is not constant. This proves that $X_r$ is a monotonically decreasing function of $r$ on the entire interval of existence. In particular, we have that
\[X_r(r)< X_r(r_0)\]
on this interval; therefore, 
\[X(r)<X(r_0)+X_r(r_0)(r-r_0)\]
for all $r>r_0$. That is, $X$ does not blow up at some finite $r$. Since $X$ and $X_r$ do not blow up for finite $r$,  the solution may be extended to the interval $(r_b,\infty)$.

We claim that $r_b\ge 0$ and as $r\to r_b^+$ the function  $X$ approaches zero and $X_r$ approaches infinity. Under our assumption that $X$ and $X_r$ are positive at the initial point $r_0$, $X$ is decreasing and $X_r$ is increasing as $r$ decreases toward the left-hand endpoint $r_b$ of its maximal interval of existence.  Suppose there is some point $p$ in the interior of this interval such that $X(p)=0$.  The function $X$ is then defined in an open interval containing $p$; therefore,  $X_{rr}(p)$ is finite. By inspection of the differential equation we must have $X_r(p)=0$. But, this is impossible because $X_r$ increases from its positive initial value. Thus, we may assume that $X>0$ on its maximal interval of existence. If $r_b<0$, then $r=0$ is an interior point of the interval of existence and again $X_r(0)=0$, in contradiction. Therefore, $r_b\ge 0$ and $X>0$ on its maximal interval of existence. Suppose that  $X$ is bounded above zero on this interval and $r_b>0$. Then, in view of equation~\eqref{eq:factB}, for $r_b<r<r_0$ and 
\[ b:=\inf_{r_b<r<r_0} \frac{1}{2} r^2 X^2(r),\] 
we have the inequality
\begin{equation}\label{eq:ineq}
-\frac{d}{dr}(X_r^2(r))<\frac{1}{b}X^2_r(r)
\end{equation}
or 
\begin{equation}\label{eq:ineq2}
\frac{d}{dr} \ln(X_r^2(r))>-\frac{1}{b}.
\end{equation}
By integrating both sides of this inequality on the interval $(r,r_0)$ and rearranging, we find that 
\begin{equation}\label{eq:ineq3}
X_r^2(r)< e^{-(r-r_0)/b} X_r^2(r_0).
\end{equation}   
In particular, $X_r$ is bounded on the interval $(r_b,r_0)$. But, since the  interval of existence was chosen to be maximal, this is impossible by the standard extension theorem for ordinary differential equations: solutions continue to exist until they reach the boundary of the domain of definition of the differential equation or they become unbounded. In effect, the solution under consideration here must continue to exist unless $X$ reaches zero, which we have excluded, or $X_r$ grows to infinity. But, using our assumption that $X$ is bounded above zero,  we have proved that $X_r$ is bounded; thus, we have reached a contradiction. It follows that $X$ approaches zero as $r$ approaches $r_b$.

We will show that $X_r$ blows up to infinity as $r$ approaches $r_b$.
By writing the differential equation in the form
\begin{equation}\label{eq:c1}
\frac{X_{rr}}{X_r}=-\frac{1}{r^2 X^2},
\end{equation}
integrating both sides from $r$ to $r_0$ and rearranging, we find that
\begin{equation}\label{eq:c2}
X_r(r) = X_r(r_0) \exp\Big(\int_r ^{r_0}\frac{1}{s^2 X^2(s)}\,ds\Big).
\end{equation}
Because $X_{rr}<0$, the graph of $X$ is concave down; therefore, this graph lies below each of
its tangent lines.
We know that $X_r$ is increasing as $r\to r_b^+$. If $X_r$ is unbounded on the interval $(r_b,r
_0)$, we have $\lim_{r\to r_b^+}X_r(r)=\infty$, as desired. On the other hand, if $X_r$ is bounded then
$\lim_{r\to r_b^+}X_r(r)=K:=\sup_{r_b<r<r_0}X_r(r)<\infty$. It follows that $X(r)<K(r-r_b)$, where $r \mapsto K(r-r_b)$ is the function giving the  limit tangent line to $X$ at $r_b$.  By inserting  the inequality
\begin{equation}\label{eq:c3}
\frac{1}{X^2(r)}>\frac{1}{K^2 (r-r_b)^2}
\end{equation}
in Eq.~\eqref{eq:c2},
it is easy to see that
\begin{equation}\label{eq:c4}
X_r(r )>  X_r(r_0) \exp\Big(\frac{1}{K^2 r_0^2}\int_r ^{r_0}\frac{1}{(s-r_b)^2}\,ds\Big)
\end{equation}
for $r_b<r<r_0$. But, the last inequality implies that $X_r(r)\to \infty$ as $r\to r_b^+$, in contradiction to the assumption that $X_r$ is bounded on $(r_b,r_0)$.

To determine the admissible solutions that allow jets, we consider the new scale-invariant functions 
\begin{equation}\label{eq:w1}
x=r^{3/2} X_r(r), \qquad y=r^{1/2} X(r) 
\end{equation}
(first found by Weishi Liu~\cite{WL})
and note that 
\begin{equation}\label{eq:w2}
\frac{dx}{dr}=\frac{x}{r} (\frac{3}{2}-\frac{1}{y^2}), \qquad \frac{dy}{dr}=\frac{1}{r} (x+\frac{1}{2} y).
\end{equation}
Thus, we have  determined a first-order system equivalent to the second-order differential Eq.~\eqref{eq:fode}. By the change of independent variable $r=e^s$, this system is made autonomous. In fact, with 
\begin{equation}\label{eq:w3}
\xi(s)=x(e^s), \qquad \eta(s)=y(e^s),
\end{equation}
system~\eqref{eq:w2} is transformed to 
the autonomous system 
\begin{equation}\label{eq:w4}
\dot \xi=\frac{\xi}{\eta^2}(\frac{3}{2}\eta^2-1), \qquad \dot\eta=\xi+\frac{1}{2} \eta,
\end{equation} 
where the overdot denotes differentiation with respect to the new independent variable $s$.

Recall that solutions are admissible if $X X_r>0$ and $X (r X_r-X)>0$. We will consider only solutions with $X>0$ for simplicity. Solutions with $X<0$ have similar properties by symmetry. 
The first condition is satisfied if both $\xi$ and $\eta$ are positive; that is, solutions that start in the (open) first quadrant of the $(\xi,\eta)$ space remain there. Because,
\begin{equation}\label{eq:w5}
X (r X_r-X)=\frac{y}{r}(x-y),
\end{equation}
a positive solution of the differential Eq.~\eqref{eq:fode} is admissible exactly when it is confined to the sector $\Sigma$ in the open first quadrant where $\xi>\eta$. 

The positive $\xi$ axis is part of the boundary of the domain of definition of system~\eqref{eq:w4}. In particular, solutions of this system starting in $\Sigma$ do not exit along this ray.  An easy calculation using system~\eqref{eq:w4} yields the inequality
\begin{equation}\label{eq:w6}
(\dot \xi -\dot \eta)\big|_{\xi=\eta}=-\frac{1}{\xi}.
\end{equation} 
It follows that $\Sigma$ is not positively invariant because
solutions that meet the upper boundary of $\Sigma$ leave this sector.

An admissible solution (that is, one that stays in $\Sigma$) allows jets exactly when $F(r)=r^2 X(r) X_r(r)-1$ vanishes at some $r>r_b$. In the new coordinates, 
\begin{equation}\label{eq:w7}
r^2 X X_r-1=xy-1.
\end{equation} 
From Eq.~\eqref{eq:w4}, we have that   
\begin{equation}\label{eq:w8}
\frac{d}{ds}(\xi\eta-1)\Big|_{\xi\eta=1}=2>0.
\end{equation}
Therefore, solutions of system~\eqref{eq:w4} starting in the sector $\Sigma$ cross the hyperbola $\xi\eta=1$ at most once. 

It is convenient to consider the change of coordinates
\begin{equation}\label{eq:w9}
u=\frac{1}{\xi}, \qquad v=\frac{\eta}{\xi},
\end{equation}
where the line $u=0$ may be viewed as the line at infinity (see~\cite{dla}). Using the new coordinates, system~\eqref{eq:w4} is transformed to
\begin{equation}\label{eq:w10}
\dot u=-\frac{u}{v^2}(\frac{3}{2}v^2-u^2), \qquad \dot v=\frac{1}{v}(v-v^2+u^2).
\end{equation}
It has the same phase portrait in the first quadrant as the system
\begin{equation}\label{eq:w11}
\dot u=u (u^2-\frac{3}{2}v^2), \qquad \dot v=v(v-v^2+u^2).
\end{equation}
The upper boundary of $\Sigma$  corresponds to the line $v=1$  in the new coordinates,  and the curve $xy=1$ (representing $F(r)=0$) corresponds to $v=u^2$.

\begin{figure}
\begin{center}
\psfig{file=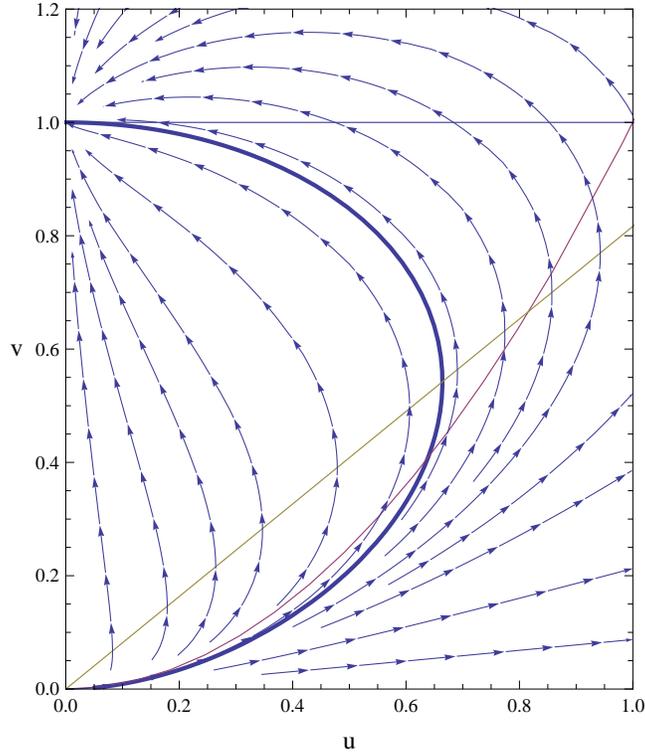, width=20pc}
\end{center}
\caption{A portion of the $(u,v)$ state space for system~\eqref{eq:w11} is shown here. We have plotted the line $v=1$, the parabola $v=u^2$, the line $v=(2/3)^{1/2} u$ (which is the vertical isocline) and the thick curve $\mathcal{Z}$ that is  an approximation of the solution that approaches the rest point at $(0,1)$ tangent to $v=1$. This solution $\mathcal{Z}$ connects the two rest points and crosses the parabola.  
\label{fig:w1}}
\end{figure}
We note that system~\eqref{eq:w11} has two rest points on the line at infinity: $(u,v)=(0,0)$ corresponding to the boundary cylinder at $r=r_b$ and $(u,v)=(0,1)$ corresponding to the axis of symmetry at $r=\infty$. The second rest point is a hyperbolic sink, and the system matrix of its linearization is diagonal. The eigenvalue of the system matrix corresponding to the vertical direction is $-3/2$ and the eigenvalue corresponding to the horizontal direction is $-1$. It follows from basic invariant manifold theory that there is an analytic solution $\mathcal{Z}$ which approaches the rest point $(0,1)$ tangent to the line $v=1$ and has the Taylor series expansion
\[ v=1-\frac{1}{2} u^2-\frac{3}{10} u^4+O(u^6).\]
Therefore, $\mathcal{Z}$ approaches the rest point $(0,1)$ from below the line $v=1$. Numerical experiments suggest that  $\mathcal{Z}$ also approaches the origin (in the backward direction of the independent variable) and it crosses the curve $v=u^2$ (see Fig.~\ref{fig:w1}).

Our numerical evidence suggests that all solutions of system~\eqref{eq:w11} starting in the first quadrant to the left of the curve $\mathcal{Z}$ correspond to admissible solutions that allow jets. It is clear that every solution starting to the left of  $\mathcal{Z}$ remains to the left; therefore, such a solution does not cross the line $v=1$. This means that corresponding solutions remain in $\Sigma$ for all time. As previously mentioned, this fact implies that the quantity $X(r)(r X_r(r)-X(r))$ is positive for all $r$ so that these solutions are admissible. Figure~\ref{fig:w1} strongly suggests that an open segment of the parabola $v=u^2$ lies to the left of $\mathcal{Z}$. It follows that all solutions starting on this portion of the parabola correspond to spacetimes that allow jets.

We note that Fig.~\ref{fig:w1} indicates the range of $\beta_{\mathtt{min}}$ defined in Eq.~\eqref{eq:100n}. Indeed, using the coordinates defined in this appendix, 
\begin{equation}\label{eq:w100}
\beta_{\mathtt{min}}=y=\eta=\frac{v}{u}
\end{equation}
at points in $(u,v)$ space on the curve $v=u^2$, which corresponds to evaluation at $r = r_J$. 
Thus, we have that 
\begin{equation}\label{eq:w101}
\beta_{\mathtt{min}}=u
\end{equation}
for those values of $u$  such that $(u,u^2)$ is in the region of admissible solutions to the left of the curve $\mathcal{Z}$ depicted in Fig.~\ref{fig:w1}. The corresponding range of $u$ is approximately the interval $(0,0.63)$. 

The nonlinear system~\eqref{eq:w11} will behave asymptotically the same as its linearization near the hyperbolic rest point $(0,1)$, which is given in the fourth quadrant by 
\[ \dot U =-\frac{3}{2}U, \qquad \dot V=- V.\] 
Its solutions lie on curves of the form \[V= -a^2 U^{2/3};\] hence, the corresponding solutions of the nonlinear system lie on curves with  asymptotic expansions of the form 
\[v=1-a^2  u^{2/3} + b u^2+O(u^{5/2}).\]
Also, since the solution  $\mathcal{Z}$  is analytic and tangent to the horizontal axis, a computation with power series can be used to show that it lies on an analytic curve of the form \[v=1-\frac{1}{2} u^2+O(u^3)\] 
near this rest point. Thus, all the admissible solutions have expansions of the form
 \[v=1-a^2 u^{2/3}-\frac{1}{2} u^2+O(u^{5/2}).\]

For $v=1$, which corresponds to the leading-order behavior of an admissible solution as $r\to \infty$, we have that $y=x$ and $X(r)=r X_r(r)$. The solutions of this differential equation have the form $X(r)=a r$, for some constant $a>0$. This suggests the asymptotic behavior of the differential Eq.~\eqref{eq:fode} as $r\to \infty$ might be obtained by studying solutions of the form  $X(r)=a r S(1/r)$  for some function $S$ such that $S(0)=1$. By inserting this relation into the original differential Eq.~\eqref{eq:fode} and using $\eta=1/r$, we find that if $S_0$ is a constant and $S$ satisfies the initial value problem 
\begin{equation}
a^2 S^2 S''-\eta^2 S'+\eta S=0,\quad S(0)=1,\quad  S'(0)=S_0,
\end{equation}
then $X(r)=a r S(1/r)$ is a solution of~\eqref{eq:fode}. We note that the differential equation is not singular at $S=1$. Therefore the initial value problem has  an analytic solution, which may be obtained by inserting a formal series representation for $S$ in powers of $\eta$ and equating coefficients. The solution has the form 
\[ S(\eta)=1-\frac{\eta^3}{6 a^2} +O(\eta^4)+S_0 \eta(1+\frac{\eta^3}{6 a^2} +O(\eta^4)).\]
It corresponds to 
\begin{equation}\label{eq:2parm}
X(r)=a r (1+\frac{S_0}{r}-\frac{1}{6 a^2r^3}+\frac{S_0}{6a^2r^4}+O(\frac{1}{r^5})).
\end{equation}
We have that $\lim_{r\to \infty} rX_r(r)-X(r)=-a S_0$.  But for admissible solutions, the quantity  $rX_r(r)-X(r)$ has the same sign as $X(r)$; thus,  we see that $-a S_0$ must be chosen to have the sign of $X(r)$. That is, under our assumption that $X(r)$ is positive, we take $S_0<0$.

While the numerical evidence is compelling, we will prove that there is an open set of admissible spacetimes that allow jets. Our argument uses new coordinates. We note first that for the basic differential Eq.~\eqref{eq:fode}, we have $X(r_b)=0$ and $X_r(r_b)=\infty$. To avoid the infinite derivative, we consider instead the inverse function $Y$, which is defined by the relations $Y(X(r))=r$ and $X(Y(s))=s$, and note that it satisfies the initial value problem
\begin{equation}\label{eq:hsi1}
s^2 Y^2Y_{ss}=Y_s^2, \qquad Y(0)=r_b, \quad Y_s(0)=0.
\end{equation} 
We define the scale-invariant quantities 
\[x(s)=s^3 Y_s(s), \qquad y(s)=s^2 Y(s)\]
and 
\[\xi(t)=x(e^{-t}), \qquad \eta(t)=y(e^{-t})\]
to obtain the corresponding first-order system
\begin{equation}\label{eq:hsi2}
\dot \xi=-\frac{\xi}{\eta^2} (3 \eta^2+\xi), \quad 
\dot \eta=-(\xi+2\eta).
\end{equation}
It is convenient to make the change of coordinates
\begin{equation}\label{eq:hsi3a}
\xi=\frac{p^2}{q}, \qquad \eta=\frac{p}{q}
\end{equation}
that transforms system~\eqref{eq:hsi2} to
\begin{equation}\label{eq:hsi3}
\dot p=p(p-q-1), \qquad \dot q=q(1+2 p -q).
\end{equation}
We note that system~\eqref{eq:hsi3} is a Lotka-Volterra system, a family of differential equations that arises in many applications such as, for instance, chemical kinetics and ecology~\cite{27}.

Recall that the admissibility conditions are $X X_r>0$ and  $X (r X_r-X)>0$. The first condition is effectively redundant. Indeed the second condition is equivalent to 
\[r\frac{ X_r(r) }{X(r)}-1>0,\]
which can only be satisfied for $X X_r>0$. 
Note that 
\[r\frac{ X_r(r) }{X(r)}=\frac{Y(s)}{s Y_s(s)}=\frac{y(s)}{x(s)}=\frac{\eta(t)}{\xi(t)}=\frac{1}{p(t)},\]
hence
\begin{equation}\label{eq:t1} p(t)=\frac{X}{r X_r}, \qquad q(t)=\frac{1}{r^2 X X_r},\end{equation}
where $t=-\ln X$. 
Thus, a solution is admissible when 
$0<p(t)<1$
for all $t$.
Similarly, an admissible solution allows jets when $r^2 X(r) X_r(r)-1$ vanishes for some $r>r_b$. This condition is satisfied for those solutions such that  $q(t)=1$  for some finite $t$, where $t=-\infty$ at the symmetry axis and $t=\infty$ at the boundary. 

\begin{figure}
\begin{center}
\psfig{file=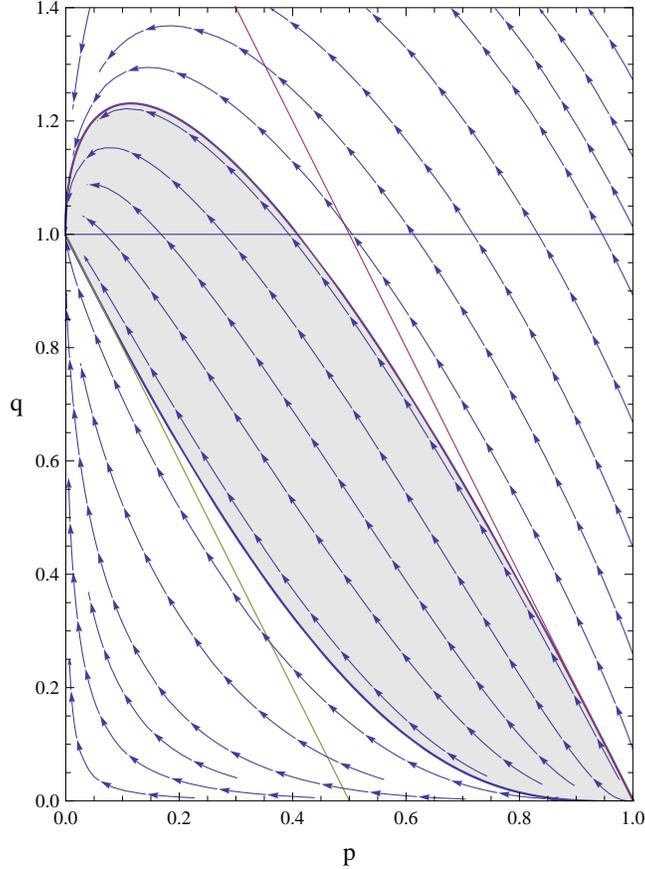, width=20pc}
\end{center}
\caption{ A portion of the phase portrait of system~\eqref{eq:hsi3}  is shown. The shaded region is bounded by the trajectory $\mathcal{J}$ that leaves the source at $(p,q)=(1,0)$ tangent to the line  $q=-2 p+2$ (which is also depicted) and the trajectory that approaches the sink at $(0,1)$ tangent to the line $q=-2 p+1$ (which is also depicted).  Solutions that correspond to trajectories that  lie in the shaded region are admissible (because their $p$ components remain in the interval $0<p<1$) and allow jets (because they cross the line $q=1$). 
\label{fig:pq}}
\end{figure}
To show that there are admissible solutions that allow jets, we analyze the phase portrait of system~\eqref{eq:hsi3}---see~Fig.~\ref{fig:pq}. There are several important properties: $(i)$ There is a hyperbolic saddle point at the origin whose stable manifold is the $p$ axis and whose unstable manifold is the $q$ axis. In particular, the axes are invariant sets. $(ii)$ There is a hyperbolic source at $(p,q)=(1,0)$, which corresponds to the axis of symmetry ($r=\infty$), and a smooth solution that approaches this rest point in negative  time tangent to the line $q=-2 p+2$ with both $p$ and $q$ components positive.
$(iii)$ There is a hyperbolic sink at $(p,q)=(0,1)$, which corresponds to the boundary ($r=r_b$), and a smooth solution that approaches this rest point in positive time tangent to the line $q=-2 p+1$ with both $p$ and $q$ components positive. Moreover, this solution approaches the sink faster than the solutions along the $q$ axis. $(iv)$ Solutions cross the segment of the line $p+q=1$ between this source and sink from below  to above. More precisely, they cross in the positive direction of the normal to the line, which has components $(1,1)$. $(v)$ Solutions cross the line $p=1$ from right to left. $(vi)$ Solutions cross the line $q=1$ from below to above.  Facts $(i)$, $(ii)$ and  $(iii)$ are proved by linearization and stable manifold theory. Facts $(iv)$, $(v)$ and $(vi)$ are proved by checking the direction of the vector field associated with system~\eqref{eq:hsi3} on the lines.  

Given the properties just enumerated, consider a solution $\mathcal{J}$ leaving the source tangent to the line  $q=-2 p+2$ and note that $\mathcal{J}$ starts above the line $p+q=1$. Hence, by $(iv)$ it stays above the line. Also by $(v)$ and $(vi)$,  $\mathcal{J}$ is confined to the triangle with corners $(1,0)$, $(1,1)$ and $(0,1)$ unless it crosses the line $q=1$. It is easy to see---by inspection or one may use the Poincar\'e-Bendixson theorem---that $\mathcal{J}$ cannot remain in the triangle for all time unless it approaches the sink. In fact,  it does approach this rest point but not until after it crosses the line $q=1$. To prove this fact requires a further analysis of the local behavior at the sink. The important point is that $\mathcal{J}$ approaches the sink above the line $q=-2 p+1$. A solution of the system approaching this \emph{hyperbolic} rest point tangent to this line---according to $(iii)$---approaches faster than the solutions approaching along the $q$ axis. It follows that all solutions except those on the trajectory approaching tangent to  $q=-2 p+1$ must approach the sink tangent to the $q$ axis. Hence, the solutions approaching above this tangent must cross the line $q=1$, as required. 

It is interesting to note here the connection between these results and Eq.~\eqref{eq:20m}, which implicitly expresses the way $X(r)$ approaches zero as $r\to r_b$ relative to the parameter $A$. The trajectory that approaches the sink at the boundary $(p,q)=(0,1)$ tangent to $q=-2p +1$ uniquely corresponds to $A=0$, see Eq.~\eqref{eq:A15}. All other trajectories must approach this sink tangent to the $q$ axis and have $A\ne 0$; while those with $A>0$ stay below $q=1$ and have no jets, the admissible solutions that allow jets all have $A<0$. As shown in Fig.~\ref{fig:pq}, there is a negative  critical value of $A$ (corresponding to trajectory $\mathcal{J}$) below which the corresponding solutions are not admissible.   

If $\mathcal{J}$ crosses the line $q=1$ at a point with coordinates $(j, 1)$, then all solutions crossing the line $q=1$ with $p$ coordinate in the interval $(0,j)$ are admissible and allow jets. It follows that $j$ is the same as the maximum possible $\beta_{\mathtt{min}}^2$, which is approximately 0.4 according to Figs.~\ref{fig:w1} and~\ref{fig:pq}. 

Let us return to the behavior of the solutions of Eq.~\eqref{eq:fode} as $r\to r_b^+$ to prove Eq.~\eqref{eq:20m}. This series representation may be found formally in several ways. To prove that it is correct seems to require several steps that are outlined here.  

In the coordinates of system~\eqref{eq:hsi3}, which are related to solutions of Eq.~\eqref{eq:fode} in display~\eqref{eq:t1}, the corresponding  problem is the asymptotic behavior near the rest point at $(p,q)=(0,1)$. For computational convenience---that is, the ability to compute at the origin with a system whose linear part is diagonal---we change coordinates via
\[P=p,\qquad Q=q-1+2 p\]
to obtain the equivalent system
\begin{equation}\label{eq:t2}
\dot P=-2 P+3 P^2-PQ,\qquad \dot Q=-Q-2 P^2+4 PQ-Q^2.
\end{equation} 
Since the eigenvalue $-2$ of the coefficient matrix of the linear part of this system is twice the other eigenvalue $-1$, normal form theory (see~\cite{arnold}) implies that there is a near-identity analytic transformation defined in a neighborhood of the origin that transforms system~\eqref{eq:t2} to 
\begin{equation}\label{eq:t3}\dot x_1=-2 x_1+\gamma x_2^2, \qquad \dot x_2=-x_2\end{equation}
for some constant $\gamma$. By performing the reduction to normal form to second order, we find that $\gamma=0$ for system~\eqref{eq:t2}. More precisely,  the near-identity change of coordinates 
\[
x_1=P-\frac{3}{2} P^2 +P Q,\qquad x_2=Q+\frac{2}{3}P^2-2 PQ+Q^2 
\]
transforms system~\eqref{eq:t2} to a system with the same linear part and a nonlinear part without quadratic terms. Stated differently,  our system is linearizable in a neighborhood of the origin; that is, there exists a near-identity analytic change of coordinates defined in a neighborhood of the origin that transforms system~\eqref{eq:t2} to~\eqref{eq:t3} with $\gamma=0$. This fact implies that system~\eqref{eq:t2} has an analytic first integral defined in a neighborhood of the origin.  More precisely, if $x_1=P+G(P,Q)$ and $x_2=Q+H(P,Q)$ is the near-identity change of coordinates (where $G$ and $H$ contain no linear terms), we may simply transform the analytic integral $x_2^2-k x_1=0$ for system~\eqref{eq:t3} with $\gamma=0$ to 
\[(Q+H(P,Q))^2= k(P+G(P,Q))\] 
to obtain the desired analytic first integral of system~\eqref{eq:t2}. Since the physical domain corresponds to $P>0$, the corresponding constant $k$ is positive. We may therefore define $k=A^2$, where $A$ can be positive or negative.  Using this fact together with the implicit function theorem, it follows that the differential equation
\[
\frac{dQ}{dP}=\frac{-Q-2 P^2+4 PQ-Q^2}{-2 P+3 P^2-PQ}
\]
has an analytic family of solutions passing through the singularity at the origin. In fact, these solutions have the form
\begin{equation}\label{eq:t4}
P=\frac{1}{A^2}(Q^2-Q^3+\frac{2 A^2+5}{2A^2} Q^4+O(Q^5)).
\end{equation}

We will determine the behavior of $X$ as $r\to r_b$. To this end, we assume that $A\ne 0$, view the integral~\eqref{eq:t4} in the abstract form 
$P=f(Q, A^2)$
and change coordinates back to $(p,q)$-coordinates with $\zeta:=q-1$ to obtain
the relation
\begin{equation}\label{eq:iB4}
p-f(\zeta+2 p,A^2)=0.
\end{equation}
Let $\mathcal{F}(p,\zeta,A)=p-f(\zeta+2 p,A^2)$. 
We have $\mathcal{F}(0,0,A)=0$ and $\mathcal{F}_p(0,0,A)=1$; hence, by the implicit function theorem, we may solve for $p$ as an analytic function of $\zeta$ and $A$ at $(p,\zeta,A)=(0,0,A)$. Let us write the result as 
\begin{equation}\label{eq:iB5}
p=g(\zeta,A^2).
\end{equation}
This relation must hold for solutions of system~\eqref{eq:hsi3} near the rest point at $(p,q)=(0,1)$. Using the second differential equation in  system~\eqref{eq:hsi3}, we have that 
\begin{equation}\label{eq:q1}
\dot\zeta=(\zeta+1)(2 g(\zeta,A^2)-\zeta)=-\zeta+O(\zeta^2).
\end{equation} 
The right-hand side of this differential equation is an analytic function of $\zeta$ defined in a neighborhood of $\zeta=0$. Of course, we may recover solutions of system~\eqref{eq:hsi3} simply as
\[p(t)=g(\zeta(t),A^2),\qquad q(t)=1+\zeta(t).\]

Using the coordinates defined in displays~\eqref{eq:hsi1} and~\eqref{eq:t1} and noting that $s=e^{-t}$, we have 
\begin{equation}\label{eq:q2}
X(r)=e^{-t}, \qquad r X^2(r)=\frac{p(t)}{q(t)}
=\frac{g(\zeta(t),A^2)}{1+\zeta(t)}.
\end{equation}

The linearization of the scalar differential Eq.~\eqref{eq:q1} has eigenvalue $-1$; therefore, this differential equation is linearizable. More precisely, there is a near-identity analytic change of coordinates 
\begin{equation}\label{eq:q7} 
\zeta=\phi+H(\phi)
\end{equation}
(defined near $\zeta=0$) that transforms Eq.~\eqref{eq:q1} to 
$\dot \phi=-\phi$, which has the general solution $\phi(t)=\phi_0 \exp(-t)$,
where $\phi_0=\phi(0)$ is an integration constant.
Thus,  Eq.~\eqref{eq:q1} has the corresponding solution 
\[\zeta(t)=\phi_0 e^{-t}+H(\phi_0 e^{-t}).\]
In view of Eq.~\eqref{eq:q2}, we have 
\begin{align}
\label{eq:q3}
\begin{split}
r X^2(r)&=\frac{g(\phi_0 X(r)+H(\phi_0 X(r)),A^2)}{1+\phi_0 X(r)+H(\phi_0 X(r))}.
\end{split}
\end{align}

We will show that it is possible to solve for $X(r)$ in this equation. 
To this end, we first determine the series representation of $H$ at the origin.
Using Eq.~\eqref{eq:q7}, we have the relation
\[\dot \zeta=(1+H_\phi(\phi,A^2))\dot \phi=-\phi (1+H_\phi(\phi,A^2)).\]
After substitution for $\zeta$ (again using Eq.~\eqref{eq:q1}), we may determine the coefficients of the power series representation of  $H$.
Using the power series in Eq.~\eqref{eq:q3}, we find the relation
\[
r X^2(r)=\frac{\phi_0^2 X^2(r)}{A^2 }+\frac{\phi_0^4 X^4(r)}{2 A^4}+\frac{\phi_0^5 X^5(r)}{3 A^4}+O(X^6(r)),
\]
which may be simplified to
\begin{equation}\label{eq:iB8}
X^2\Big(\frac{\phi_0^4 }{2 A^4}+\frac{\phi_0^5 }{3 A^4}X\Big)+O(X^4)=r-\frac{\phi_0^2 }{A^2 }.
\end{equation}
Since $X(r_b)=0$, it follows that 
$
\phi_0^2=A^2 r_b.
$
Thus, $\phi_0=\pm A r_b^{1/2}$; however, for the sake of consistency with Eq.~\eqref{eq:20m}, we choose  $\phi_0=- A r_b^{1/2}$. Thus, Eq.~\eqref{eq:iB8} becomes 
\begin{equation}\label{eq:iB9}
X^2\Big(\frac{r_b^2 }{2}-\frac{A r_b^{5/2} }{3}X\Big)+O(X^4)=r-r_b.
\end{equation}

To solve for $X(r)$ in Eq.~\eqref{eq:iB9}, define 
\[\mathcal{G}(X,r)=r-r_b-X^2\Big(\frac{r_b^2 }{2}-\frac{A r_b^{5/2} }{3}X\Big)+O(X^4)\]
and note that 
\[
\mathcal{G}(0,r_b)=0,\quad \frac{\partial\mathcal{G}}{\partial X}(0,r_b)=0,\quad 
\frac{\partial^2\mathcal{G}}{\partial X^2}(0,r_b)=-r_b^2.
\]
Under the assumption that $r_b\ne 0$, a condition that holds for admissible solutions that allow jets, the Weierstrass preparation theorem (see, for example,~\cite{ccc}) implies that
\begin{equation}\label{eq:t5}
\mathcal{G}(X,r)=(a(r)+b(r)X +X^2)U(X,r),
\end{equation}
where the functions $a$, $b$ and $U$ are analytic in a neighborhood of $X=0$ and $r=r_b$ and $U(0,r_b)\ne 0$. 
 It follows that the left-hand side of Eq.~\eqref{eq:t5} is zero for $(X,r)$ near $(0,r_b)$ exactly when $a+b X +X^2=0$ or 
\begin{equation}\label{eq:iB2}
X=\frac{1}{2} (-b(r)\pm (b^2(r)-4 a(r))^{1/2}).
\end{equation}

We return to Eq.~\eqref{eq:t5}, expand both sides in power series of both variables about the point $X=0$ and $r=r_b$ and equate coefficients to get 
\begin{align}
\label{eq:t6b}
\begin{split}
a(r)&=-\frac{2}{r_b^2}(r-r_b)-\frac{2 (9-A^2)}{9 r_b^3}(r-r_b)^2+O((r-r_b)^3),\\
b(r)&=-\frac{4 A}{3 r_b^{3/2}}(r-r_b)-\frac{4 A (9-A^2)}{135 r^{5/2}}(r-r_b)^2+O((r-r_b)^3).
\end{split}
\end{align}
Substitution of these results in Eq.~\eqref{eq:iB2} results in
\begin{equation}
\label{eq:t6bb}
\begin{split}
X(r)= {}&\frac{2^{1/2}}{r_b} (r-r_b)^{1/2}+\frac{2}{3}\frac{ A}{ r_b^{3/2}}(r-r_b)+
\frac{1}{9}\frac{(9+A^2)}{2^{1/2}r_b^2}(r-r_b)^{3/2}\\
&+\frac{2}{135}\frac{ A (9-A^2)}{r_b^{5/2}}(r-r_b)^{2} +O((r-r_b)^{5/2}).
\end{split}
\end{equation}
This expansion is valid for fixed $A\ne 0$ in case $|r-r_b|$ is sufficiently small. Moreover, $X(r)$ is an analytic function of $(r-r_b)^{1/2}$ near $r=r_b$. Using this fact, the form of the representation~\eqref{eq:t6bb} may be recovered in the usual manner by substitution of an appropriate series with undetermined coefficients in differential Eq.~\eqref{eq:fode}.

Using the expansion~\eqref{eq:t6bb}, we  have 
\begin{align} 
\label{eq:iB10}
\begin{split}
F  =  r^2 X(r) X_r(r)-1
  = {}& \Big(\frac{2}{r_b}\Big)^{1/2} A (r-r_b)^{1/2} + \frac{2(6+A^2)}{3 r_b} (r-r_b)+\frac{1}{9}\frac{A(57 +A^2)}{2^{1/2} r_b^{3/2}} (r-r_b)^{3/2}\\
&  + \frac{3105+954 A^2-11 A^4}{540 r_b^2}(r-r_b)^{2}+O((r-r_b)^{5/2}).
\end{split}
\end{align}

\section{Time Reversal}\label{appen:C}
This paper has mainly treated geodesic motion in the gravitational field given by metric~\eqref{eq:8}, which was obtained from metric~\eqref{eq:7} by time reversal. The purpose of this appendix is to study the geodesics of metric~\eqref{eq:7}  with  $X(r)$      given by Eq.~\eqref{eq:fode}. As Eq.~\eqref{eq:fode} is invariant under  $X\mapsto -X$, we can, with no loss in generality, replace   $X$    by    $ -X$      in metric~\eqref{eq:7}; that is,
\begin{equation}\label{eq:D1}
-ds^2=e^{-{t}} \frac{X_r}{X}(-X^2 d{t}^2+\frac{1}{r^3} dr^2) +\frac{e^{-{t}}}{ r}(- X d{t}+d\phi)^2+e^{{t}} dz^2.
\end{equation}
We are interested in the geodesics of Eq.~\eqref{eq:D1}. If in metric~\eqref{eq:8}  we  change $\exp(t)$  to  $\exp(-t)$  and vice versa, we get metric~\eqref{eq:D1}. The same transformation turns out to be valid for the main equations of this paper that we need for this appendix. That is, in the analysis of metric~\eqref{eq:D1}, we find it convenient to deal exclusively with the  $X>0$  branch of the admissible solutions of Eq.~\eqref{eq:fode}; then, the same equations essentially follow---such as the expressions for the invariants  $\mathcal{I}_1$ and  $\mathcal{I}_2$ and  the equations of motion for the special timelike and null geodesics---except that    $\exp(\pm t)$             must be everywhere replaced by  $\exp(\mp t)$. Consider, for instance, the special timelike geodesics given by
\begin{align}\label{eq:D2}
\begin{split}
\frac{dt}{d\tau} &=r_J  (e^{t}+C_z^2)^{1/2},\\
\frac{d\phi}{d\tau}&=r_J X(r_J ) (e^{t}+C_z^2)^{1/2},\\
\frac{dz}{d\tau}&=C_z e^{-t}.
\end{split}    
\end{align} 
The solutions are
\begin{align}\label{eq:D3}
\begin{split}
{\kappa^-}(t)&= \sinh [-\frac{1}{2}|C_z| r_ {\mathcal{J}}(\tau-\tau_0)+\sinh^{-1}({\kappa^-_0})],\\
\phi&=\phi_0+X(r_J)(t-t_0),\\
z(t)&=z_0+K^-(t_0)-K^-(t).
\end{split}    
\end{align}
Here,
\begin{align}\label{eq:D4}
\begin{split}
{\kappa^-}(t)&=|C_z|e^{-t/2},\\
K^-(t)&=\frac{C_z}{r_J|C_z|^3}
\{ \kappa^-(1+{\kappa^-}^2)^{1/2}-\ln [{\kappa^-}+(1+{\kappa^-}^2)^{1/2}]\},
\end{split}    
\end{align}
and ${\kappa^-}(t_0)=\kappa^-_0$. We note that as $t\to \infty$, $\kappa^-\to 0$, $\tau\to \tau_{\mathtt{end}}$ and $z\to z_{\mathtt{end}}$, 
where
\begin{align}\label{eq:D5}
\begin{split}
\tau_{\mathtt{end}}&=\frac{2}{r_J|C_z|}\sinh^{-1}(\kappa^-_0),\\
z_{\mathtt{end}}&=z_0+K^-(t_0).
\end{split}    
\end{align}
It follows that the special timelike geodesics are \emph{incomplete} in this case~\cite{n22}. On the other hand, for metric~\eqref{eq:D1},
\begin{equation}\label{eq:D6}
\mathcal{I}_1(r_J)=4 e^{2t }r_J^4,\qquad
\mathcal{I}_2(r_J)=0,
\end{equation}
so that as            $t\to\infty$,              $ \mathcal{I}_1\to \infty$. More generally, $\mathcal{I}_1$ and $\mathcal{I}_2$ are proportional to $\exp(2 t)$ and $\exp(3t)$, respectively, in the case of metric~\eqref{eq:D1}. This is consistent with the incompleteness of the special timelike geodesics for            $t\to \infty$, since the whole spacetime region becomes singular in this limit. Let us note, however, that the special null geodesics are complete, since
\begin{equation}\label{eq:D7}
\begin{split}
t-t_0 &= r_J | \hat C_z| (\zeta-\zeta_0), \\
\phi-\phi_0 &= r_JX( r_J)| \hat C_z| (\zeta-\zeta_0),\\
z-z_0 &= \frac{1}{  r_J }\frac{ \hat C_z}{ | \hat C_z| }(e^{-t_0}-e^{-t} ),
\end{split}
\end{equation}
so that $z(t=\infty)=z_0+\hat {C}_z \exp(-t_0)/(r_J|\hat C_z|)$.

We conclude that the special (timelike and null) geodesics in this case simply wrap infinitely around the axis of symmetry, while $z$ remains forever finite.

\acknowledgments{CC was supported in part by the NSF grant DMS 0604331.}


\begin{thebibliography}{xxxx}

\bibitem{1} B. Mashhoon, J. C. McClune and H. Quevedo, Class. Quantum Grav. \textbf{17},  533  (2000) [arXiv: gr-qc/9609017].
\bibitem{n2} $<$http://apod.gsfc.nasa.gov/apod/ap100314.html$>$.

\bibitem{f} F.~de Felice and O.~Zanotti, Gen. Rel. Grav. \textbf{32}, 1449 (2000).

\bibitem{2} B. Punsly, \emph{Black Hole Gravitohydromagnetics},  2nd. ed.  (Springer-Verlag, Berlin, 2008). 

\bibitem{3} B. Mashhoon and  J. C. McClune, Mon. Not. R. Astron. Soc. {\bf 262}, 881 (1993).

\bibitem{4} C. Chicone and B. Mashhoon, Class. Quantum Grav.
  \textbf{19}, 4231 (2002).

\bibitem{5} C. Chicone, B. Mashhoon and B. Punsly,
  Int. J. Mod. Phys. D \textbf{13}, 945 (2004); Phys. Lett. A
  \textbf{343}, 1 (2005).

\bibitem{6} C. Chicone and B. Mashhoon, Class. Quantum
  Grav. \textbf{21}, L139 (2004); Class. Quantum Grav. \textbf{22},
  195 (2005); Ann. Phys. (Berlin) \textbf{14}, 290 (2005); Astron. \&
  Astrophys. \textbf{437}, L39 (2005); Ann. Phys. (Berlin)
  \textbf{14}, 751 (2005); Class. Quantum Grav. \textbf{23}, 4021 (2006). 

\bibitem{7} Y. Kojima and K. Takami, Class. Quantum Grav. \textbf{23},
 609 (2006)  [arXiv: gr-qc/0509084].
 \bibitem{9} C. Chicone and B. Mashhoon, Phys. Rev. D \textbf{74}, 064019 (2006). 
 \bibitem{13}  C.W.F. Everitt  \emph{et al.}, {Space Sci. Rev.} \textbf{148}, 53 (2009).
 
 \bibitem{16} J.L. Synge,  \emph{Relativity: The General Theory} (North-Holland, Amsterdam, 1960). 
 
 \bibitem{14}  H. Stephani, D. Kramer, M. MacCallum, C. Hoenselaers and E. Herlt,
\emph{Exact Solutions of Einstein's Field Equations}, 2nd ed. (Cambridge
University Press, Cambridge, 2003), ch. 22.

\bibitem{11n} W.B. Bonnor, J.B. Griffiths and M.A.H. MacCallum, Gen. Rel. Grav. \textbf{26}, 687 (1994).

\bibitem{15} L. Wylleman and N. Van den Bergh, {Phys. Rev. D} \textbf{74}, 084001 (2006).

\bibitem{13n} R. Opher, N.O. Santos and A. Wang, J. Math. Phys. \textbf{37}, 1982 (1996).

\bibitem{14n} L.M. Berkovich, Symmetry in Nonlinear Math. Phys. \textbf{1}, 155 (1997);  H. Goenner and P. Havas, J. Math. Phys. \textbf{41}, 7029 (2000).

\bibitem{8} A. Lichnerowicz, \emph{Th\'eories Relativistes de la Gravitation
  et de l'\'Electromagn\'etisme} (Masson, Paris, 1955).

\bibitem{ccc} C. Chicone, \emph{Ordinary Differential Equations with Applications}, 2nd ed. (Springer-Verlag, New York, 2006).
\bibitem{19} M.A.H. MacCallum, Gen. Rel. Grav. \textbf{30}, 131 (1998); J. Carot, J.M.M. Senovilla and R. Vera, Class. Quantum Grav. \textbf{16}, 3025 (1999). 

\bibitem{11} B. Mashhoon, ``On the gravitational analogue of Larmor's theorem", {Phys. Lett. A}  \textbf{173}, 347 (1993).

\bibitem{12}  B. Mashhoon,``Gravitoelectromagnetism: A Brief Review",  in 
\emph{The Measurement of Gravitomagnetism: A Challenging Enterprise}, edited by L.
Iorio (NOVA Science, Hauppauge, New York, 2007), pp. 29--39 [arXiv: gr-qc/0311030].

\bibitem{17} B.~Mashhoon and N.O.~Santos, Ann. Phys. (Berlin) {\bf 9}, 49 (2000) [arXiv: gr-gc/9807063]; B.~Mashhoon, F.~Gronwald and H.I.M. Lichtenegger, Lect. Notes Phys. \textbf{562}, 83 (2001)[arXiv:gr-qc/9912027].

\bibitem{nn22}  A. E. Broderick and A. Loeb, Astrophys. J. \textbf{703}, L104 (2009) [arXiv:0908.2999].

\bibitem{nn23} F.M. Rieger, Chin. J. Astron. Astrophys. \textbf{5S}, 305 (2005); X.Y. Hong \emph{et al.}, New Astron. Rev. \textbf{43}, 699 (1999).

\bibitem{23} B. Mashhoon, Ann. Phys. (N.Y.)  \textbf{89}, 254 (1975).

\bibitem{vn24} V. Petrosian and A.M. Bykov, Space Sci. Rev. \textbf{134}, 207 (2008) [ arXiv:0801.0923].

\bibitem{WL} W. Liu, Private Communication.

\bibitem{dla} F. Dumortier, J. Llibre and J.C. Art\'es, \emph{Qualitative Theory of Planar Differential Systems} (Springer, Berlin, 2006).
\bibitem{27} J.D. Murray, \emph{Mathematical Biology}, 3rd ed. (Springer, Berlin, 2002).
\bibitem{arnold} V.I. Arnold, \emph{Geometrical Methods in the Theory of Ordinary Differential Equations} (Springer, Berlin, 1983).
\bibitem{n22}
S.W. Hawking and G.F.R. Ellis, \emph{The Large Scale Structure of Space-Time} (Cambridge University Press, Cambridge, 1973).
\end{thebibliography}
\end{document}